%% file: neurips_2024.tex
\documentclass{article}

\PassOptionsToPackage{square,comma,numbers,sort&compress}{natbib}
\usepackage[final]{neurips_2024}

\usepackage{notoccite}

\usepackage[utf8]{inputenc} 
\usepackage[T1]{fontenc}    
\usepackage{hyperref}       
\usepackage{url}            
\usepackage{booktabs}       
\usepackage{amsfonts}       
\usepackage{nicefrac}       
\usepackage{microtype}      
\usepackage{xcolor}         
\usepackage{caption}
\usepackage{subcaption}
\usepackage{amssymb}
\usepackage{import}
\usepackage{algorithm,algorithmicx,algpseudocode}
\usepackage{multirow}
\usepackage{multicol}
\usepackage{tablefootnote}
\usepackage{pifont}
\usepackage{amsmath}
\usepackage{amsthm} 
\newcommand{\cmark}{\ding{51}}%
\newcommand{\xmark}{\ding{55}}%
\usepackage{wrapfig}
\usepackage{graphicx}
\usepackage{wrapfig}
\title{SongCreator: Lyrics-based Universal Song Generation}
\author{
  Shun Lei$^{1}$, Yixuan Zhou$^{1\dagger}$, Boshi Tang$^{1}$, Max W. Y. Lam$^{2}$, Feng Liu$^{2}$, Hangyu Liu$^{2}$, 
  \\ \textbf{Jingcheng Wu$^{2}$, Shiyin Kang$^{2}$, Zhiyong Wu$^{1,3}$\thanks{Corresponding author, $^{\dagger}$ Equal contribution.}, Helen Meng$^{3}$}
    \\
  $^1$ Shenzhen International Graduate School, Tsinghua University, Shenzhen
  \\ $^2$ Independent Researcher \\
  $^3$ The Chinese University of Hong Kong, Hong Kong SAR\\
  \texttt{\{leis21, yx-zhou23\}$@$mails.tsinghua.edu.cn, zywu$@$sz.tsinghua.edu.cn} \\
}

\begin{document}

\maketitle

\input{content/abstract}
\input{content/introduction}
\input{content/relatedwork}
\input{content/method}
\input{content/experiments}
\input{content/discussion}

\section*{Acknowledgements} This work is supported by National Natural Science Foundation of China (62076144) and Shenzhen Science and Technology Program (WDZC20220816140515001, JCYJ20220818101014030).

\bibliographystyle{unsrtnat}
\bibliography{neurips_2024}

\newpage
\import{./content}{appendix.tex}

\end{document}

%% file: content/abstract.tex
\begin{abstract}
  Music is an integral part of human culture, embodying human intelligence and creativity, of which songs compose an essential part.
  While various aspects of song generation have been explored by previous works, such as singing voice, vocal composition and instrumental arrangement, etc., generating songs with both vocals and accompaniment given lyrics remains a significant challenge, hindering the application of music generation models in the real world.
  In this light, we propose SongCreator, a song-generation system designed to tackle this challenge. 
  The model features two novel designs: a meticulously designed dual-sequence language model (DSLM) to capture the information of vocals and accompaniment for song generation, and a series of attention mask strategies for DSLM, which allows our model to understand, generate and edit songs, making it suitable for various song-related generation tasks by utilizing specific attention masks.
  Extensive experiments demonstrate the effectiveness of SongCreator by achieving state-of-the-art or competitive performances on all eight tasks. 
  Notably, it surpasses previous works by a large margin in lyrics-to-song and lyrics-to-vocals. 
  Additionally, it is able to independently control the acoustic conditions of the vocals and accompaniment in the generated song through different audio prompts, exhibiting its potential applicability.
  Our samples are available at \href{https://thuhcsi.github.io/SongCreator/}{https://thuhcsi.github.io/SongCreator/}.
\end{abstract}

%% file: content/introduction.tex
\section{Introduction}
Music is an integral part of human culture, embodying human intelligence and creativity. 
Songs combining vocals and accompaniment compose an essential part of it, whose generation has been a hotspot in both academia and industry in recent years.
Although with the rapid advancements in generative models, communities have witnessed the applications of Artificial Intelligence Generated Content (AIGC) models in the generation of texts \cite{lamda, touvron2023llama, achiam2023gpt}, images \cite{esser2021taming,yu2022scaling,yu2022vectorquantized} and speeches \cite{fastspeech,vits, valle, naturalspeech,megatts}, it still remains a big question whether we can replicate the successes in song generation, which demands coordination among various complex elements such as instruments, rhythm, melody and vocals. 
Currently creating high-level songs with both vocals and accompaniment still requires substantial human effort in composition, instrument arrangement, singing, and so on, a process requiring a great deal of time and expertise. 
Lyrics-to-song generative models could lower the barrier to entry for novices and improve the workflow of experienced artists.

Previous works mostly explored specific aspects of song generation, as listed in Table \ref{tab:previous-works}. Although they exhibit abilities in vocal composition, instrumental arrangement and harmonious generation, none of them is able to combine these three for high-quality lyrics-to-song generation. To this end, Jukebox \cite{dhariwal2020jukebox} can be seen as the first and only attempt from published literature so far to simultaneously generate vocals and accompaniment in a song from lyrics using a single model. However, it exhibits two major limitations.
Firstly, this approach treats the combination of vocals and accompaniment as an entity.
While the design facilitates the generation of songs, it ignores the mutual influence between vocals and accompaniment, resulting in vocals that sound unnatural and a lack of musicality in both the melody and accompaniment, and inhibiting the independent controllability of the generated vocals and accompaniment.
Secondly, it is confined to performing specific tasks of lyrics-to-song generation, which restricts the broader application of song generation models in complex musical scenarios, including the generation of vocals or instrumental music, as well as universal song generation tasks such as song editing and accompaniment-to-vocal generation. 
Recently, while the industry has seen the emergence of song generation tools like Suno \cite{suno} and Udio \cite{udio}, neither has disclosed their methodologies nor has expanded into universal song generation tasks.

\begin{table}
  \scriptsize
  \caption{A comparison of song generation with related tasks in the literature. We use
   \textbf{Composition} to denote whether the model can complete vocal composition, \textbf{Arrangement} to denote whether the model can arrange the instrumental accompaniment, and \textbf{Harmony} to denote whether vocals and accompaniment sound harmonious and pleasant together.}
  \label{tab:previous-works}
  \centering
  \begin{tabular}{lccccc}
    \toprule
    \textbf{Tasks} & \textbf{Inputs} & \textbf{Outputs} & \textbf{Composition} & \textbf{Arrangement} & \textbf{Harmony}\\
    \midrule
    Singing Voice Synthesis \cite{chen2020hifisinger,huang2022singgan, zhang2022visinger,hong2023unisinger, liu2022diffsinger, hwang2023hiddensinger} & Scores & Vocals & \xmark & \xmark & \xmark\\
    SongComposer \cite{ding2024songcomposer} & Lyrics & Vocals & \cmark & \xmark & \xmark\\
    Text-to-Music \cite{musicgen, melody, mousai, musiclm} & Text Description &  Music & \xmark & \cmark & \xmark\\
    Accompaniment Generation \cite{grachten2020bassnet,wu2022jukedrummer,yeh2021automatic,donahue2023singsong, zhiqing2024text} & Vocals & Music & \xmark & \cmark & \cmark\\
    \midrule
    \textbf{Song Generation}  & Lyrics & Song & \cmark & \cmark & \cmark\\
    \bottomrule
  \end{tabular}
\end{table}

In this work, we introduce SongCreator, a system designed to generate high-quality songs with harmoniously coordinated vocals and accompaniment based on lyrics.
It is worth mentioning that by learning composition and arrangement abilities during training, SongCreator can also be applied to universal song generation tasks, as shown in Appendix \ref{sec:tasks}, including (but not limited to) lyrics-to-vocal, accompaniment-to-song and song editing.
Formalized as a combination of a language model (LM) and a latent diffusion model (LDM) \cite{ldm}, SongCreator features a novel dual-sequence language model (DSLM), which utilizes two decoders to separately model vocals and accompaniment information, and employs a dynamic bidirectional cross-attention module to capture the influences between these two sequences.
This approach treats vocals and accompaniment within a song as separate but interrelated sequences, effectively reducing their mutual influence during training.
Additionally, inspired by UniLM \cite{dong2019unilm} and GLM \cite{du2022glm}, we design a series of attention mask strategies for DSLM, which enables SongCreator to complete song generation tasks of various forms, such as editing, understanding and generation in a unified manner.
Our contributions can be summarized as follows:
\begin{itemize}
    \item We propose a novel dual-sequence language model for song generation. Compared to previous ones, it not only emphasizes the respective quality of vocals and accompaniment, but also learns their mutual influences to coordinate them into harmonious songs, greatly enhancing the quality of generations.
    \item We propose a series of attention mask strategies for song generation, which endows our model with the ability to unify song generation tasks of various forms, such as lyrics-to-song, accompaniment-to-song and song editing. It also makes multi-task training feasible for SongCreator, which underlies its versatile generation ability.
    \item On top of the mechanisms above, we propose a versatile system for song generation. It can be readily applied to lyrics-based vocals/song generation, or even editing. 
    It also supports universal conditioning and generation: given any one of the vocals or accompaniment as a condition, SongCreator is able to generate the other. Moreover, SongCreator is able to generate songs with separate audio prompts for vocals and accompaniment.
    \item 
    We conduct extensive experiments to demonstrate the abilities of our system in the eight tasks shown in Appendix \ref{sec:tasks}.
    Ablation experiments justify the effectiveness of our designs.
\end{itemize}

%% file: content/relatedwork.tex
\section{Related Work}
\paragraph{Singing voice synthesis} 
Singing Voice Synthesis (SVS) \cite{chen2020hifisinger,huang2022singgan, zhang2022visinger,hong2023unisinger, liu2022diffsinger, hwang2023hiddensinger} aims at synthesize vocals given scores, has made great progress in recent years. 
Several works attempt to adopt transformer models \cite{chen2020hifisinger}, generative adversarial networks \cite{huang2022singgan} and conditional variational autoencoder \cite{zhang2022visinger, hong2023unisinger} for SVS. 
Recently, research \cite{liu2022diffsinger, hwang2023hiddensinger} focuses on enhancing the quality of synthesized vocals through diffusion models, demonstrating state-of-the-art (SOTA) performance.
Similarly, SongCreator also employs a diffusion model to improve the quality of synthesized songs.
However, compared to traditional SVS methods that require additional scores composed by humans, SongCreator facilitates composing and arranging songs directly from lyrics and generates complete songs with accompaniment.

\paragraph{Music generation}
Music generation has been long studied under various setups.
Early efforts \cite{dong2018musegan,wei23c_interspeech} primarily focus on generating symbolic music, which is confined to fixed instrumental timbres and lacks expressiveness.
Several works \cite{dhariwal2020jukebox,audiolm, musiclm, musicgen} have achieved text-to-music generation by tokenizing music into discrete sequences that can be further processed by language models (LMs) \cite{achiam2023gpt, touvron2023llama}.
Singsong \cite{donahue2023singsong} and Melodist \cite{zhiqing2024text} follow a similar approach for accompaniment generation.
Diffusion models \cite{sohl2015deep, ddpm, vdm}, as another competitive class of generative models, have also delivered impressive results in music generation.
Many emerging methods \cite{chen2024musicldm, riffusion, noise2music, mousai} use latent diffusion model (LDM) to generate high-quality and high-fidelity music.
Recently, MeLoDy \cite{melody} and AudioLDM 2 \cite{liu2023audioldm} introduce a novel solution by combining the advantages of LMs and LDM, demonstrating SOTA performances with high fidelity and musicality.
However, these methods are designed for generating non-vocal music and limited to specific task such as text-to-music or vocal-to-accompaniment.
By leveraging the DSLM, SongCreator can effectively model songs that include both vocals and accompaniment, and it can also extend to various song generation tasks.
\paragraph{Speech editing and synthesis}
Speech editing requires to alter a segment within a speech to match the target transcript.
Early methods \cite{tan2021editspeech, wang2022campnet, bai20223, borsos22_interspeech, jiang2023fluentspeech} utilized the surrounding speech context as a condition, enabling models to generate the masked segment. 
Subsequently, several works \cite{megatts,du2024unicats,yin22_interspeech, yang2023uniaudio} attempted to establish a unified model for both speech editing and text-to-speech (TTS).
However, despite their impressive achievements, these efforts are restricted to handing clean signals only, and required duration information for each phoneme.
It restricts the applicability of such methods in song or vocal editing.
Recently, the advancement of LMs significantly promoted progress in speech generation, particularly in zero-shot TTS \cite{valle, speartts, zhang2023speak} and speech editing \cite{le2024voicebox, wang2023speechx, peng2024voicecraft}.
Different from these works that only focus on speech, to our knowledge, we are the first to implement song and vocal editing. 
Additionally, through a serious of attention mask strategies, our proposed SongCreator provides a general solution that enables a single system to handle multiple tasks in song generation.

%% file: content/method.tex
\section{Method} \label{sec:method}
\subsection{Overview} \label{sec:overview}
\begin{figure}
     \centering
     \includegraphics[width=\textwidth]{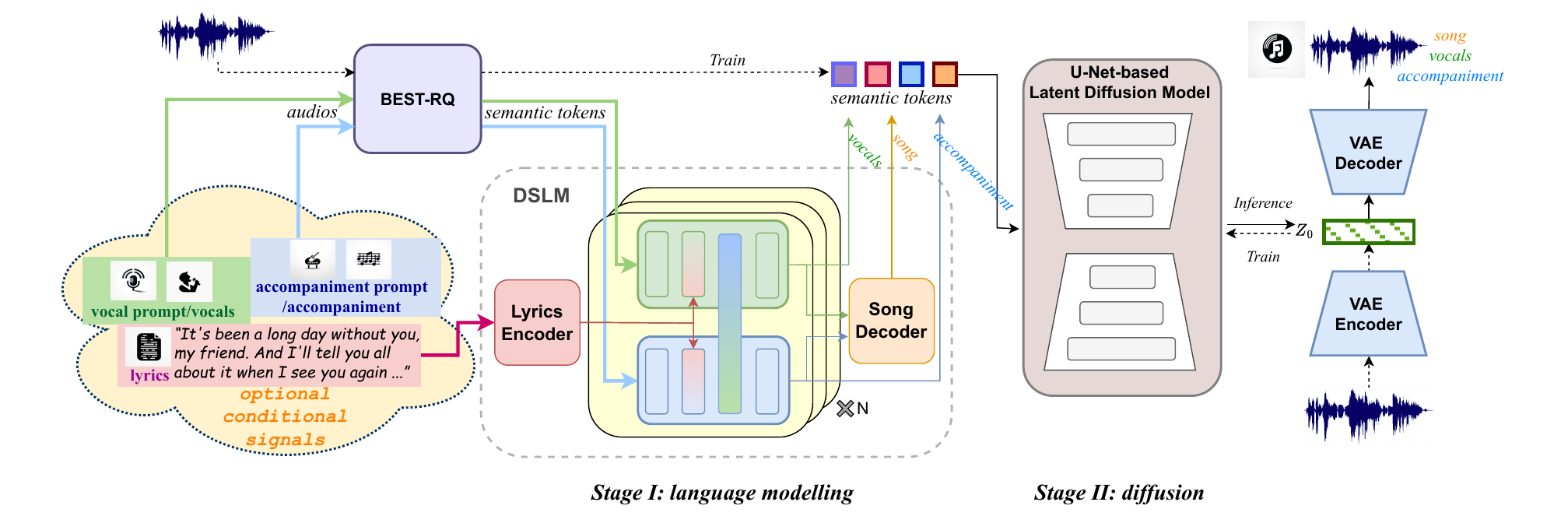}
     \caption{The overview of SongCreator. The BEST-RQ tokens is a proxy that bridges the DSLM and the latent diffusion model.}
     \label{fig:overall}
\end{figure}
Let $\mathbf{x}\in\mathcal{X}$ represent a song audio.
A song generation process can be defined as $f: \mathcal{C}\mapsto \mathcal{X}$, where $\mathcal{C}$ is the set of conditioning signals.
In this work, we consider a flexibly conditioned song generation system $f$ with $\mathbf{C}\in\mathcal{C}$, accepting a variety of optional inputs including lyrics, vocal prompt, accompaniment prompt, pre-determined vocal track and pre-determined accompaniment track.
The high flexibility of conditions empowers the controllability of our model, so that different elements within the generated songs can be customized as needed.

However, end-to-end generating a high-fidelity song $\mathbf{x}$ from $\mathbf{C}$ with a neural network $f$ remains challenging to date. 
In the same spirit as previous works \cite{audiolm,musiclm,melody}, we introduce a language-alike discrete sequence (a.k.a., \emph{semantic tokens}), denoted as $\mathbf{S}=({S}_1, \ldots, S_N)$, to capture significant structural information in song and to embody LMs as the ``brain'' of our system for writing songs.

As illustrated in Figure \ref{fig:overall}, to obtain the semantic tokens, we train a BEST-RQ \cite{bestrq} on an unlabeled dataset containing songs, vocals and music, and conduct vector quantization over its intermediate hidden representations.
These tokens encapsulate sufficient semantic and acoustic details that are necessary for reconstructing $\mathbf{x}$.
With such a purpose, an LDM, consisting of a VAE and a diffusion model, is trained to decode the semantic tokens into high-quality song audio, in a way similar to \cite{melody}. Since both BEST-RQ and LDM were trained and re-produced with open-source implementations, their respective details are beyond the focus of this paper, and are described in Appendix \ref{sec:bestrq}-\ref{sec:ldm}.

To predict the semantic tokens $\mathbf{S}$ given $\mathbf{C}$, we designed a novel form of LM -- dual-sequence language model (DSLM) for multi-condition song generation, as illustrated in Figure \ref{fig:dslm}. 
Specifically, DSLM includes three decoders, respectively adopting semantic tokens of vocals (i.e., $\mathbf{S}_v\in\mathcal{S}_v$), accompaniment (i.e., $\mathbf{S}_a\in\mathcal{S}_a$), and song (i.e., $\mathbf{S}_s\in\mathcal{S}_s$) as the prediction targets. Mathematically, we define $\text{DSLM}: \mathcal{C} \mapsto \mathcal{S}_s\times \mathcal{S}_v\times \mathcal{S}_a$.
By applying an off-the-shelf source separation algorithm to a large corpus of songs with lyrics, a large volume of paired data can be manufactured for the multi-target generation task of interest.

The remainder of this section presents the main contribution of this paper -- DSLM, and the attention mask strategies for DSLM.
\subsection{Dual-sequence language model}
\begin{figure}
     \centering
     \includegraphics[width=\textwidth]{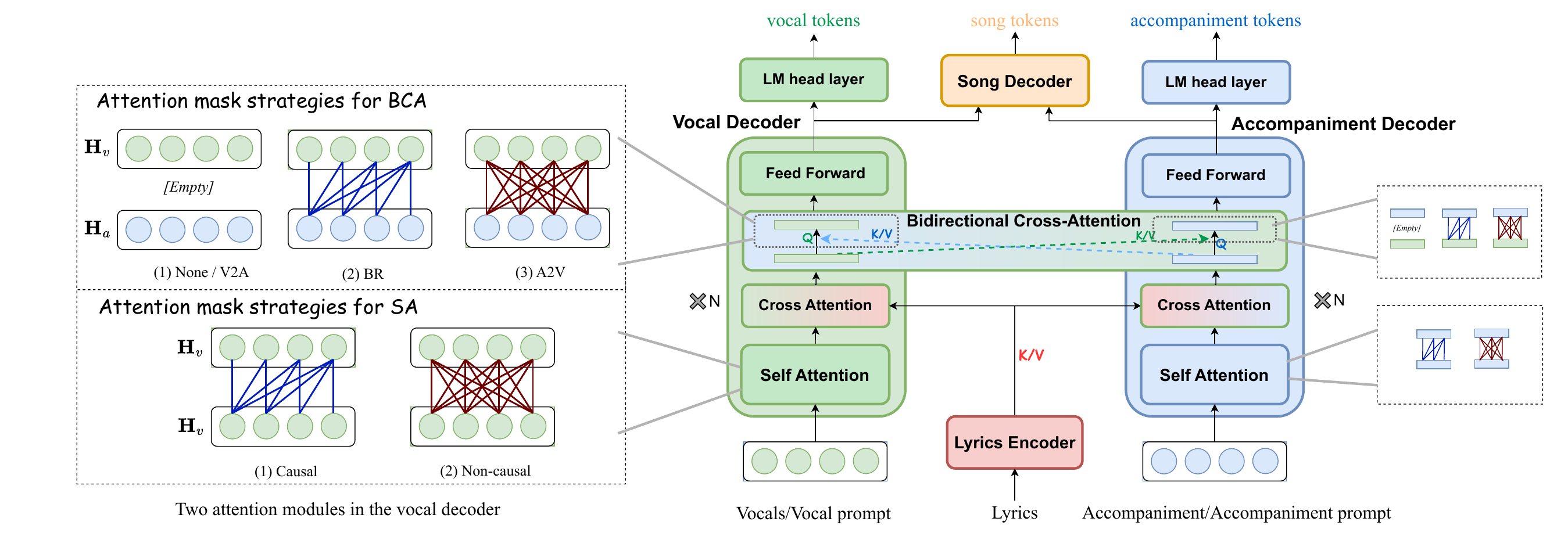}
     \caption{
     The overview of DSLM with the attention mask strategies. The DSLM can utilize specific attention mask strategy to achieve different song generation tasks.
     We illustrate multiple attention mask strategies of what
     each vocal token's representation attend to in both self-attention and bidirectional cross-attention. 
     Attention mask strategies in the accompaniment decoder are similar.
     }
     \label{fig:dslm}
\end{figure}
Formally speaking, the proposed dual-sequence language model (DSLM) is tasked with the generation of $(\mathbf{S}_s, \mathbf{S}_v, \mathbf{S}_a)$ given $\mathbf{C}$. 
An overview of the proposed architecture is presented in Figure \ref{fig:dslm}.
Concerning the quadratic complexity of Transformer with respect to sequence length, instead of processing the concatenated sequences of multiple target sequences token-by-token as in \cite{donahue2023singsong}, in DSLM we utilize different decoders to model the semantic tokens of vocals $\mathbf{S}_v$ and accompaniment $\mathbf{S}_a$ and harmoniously combine them to generate the semantic tokens of song $\mathbf{S}_s$.

The proposed DSLM consists of a lyrics encoder, two decoders (one for vocals and one for accompaniment) inter-connected through a bidirectional cross-attention module, and a final song decoder. The lyrics encoder is built upon a stack of Transformer encoder layers, which, as a architecture widely adopted in speech synthesis \cite{fastspeech,naturalspeech}, extracts critical information related to the pronunciation of the lyrics $\mathbf{C}_\text{lyrics}$. On the other hand, the vocal decoder and accompaniment decoder are together composed of multiple DSLM blocks.
Each DSLM block is composed of a self-attention (SA) layer, a cross-attention (CA) layer, a bidirectional cross-attention (BCA) layer and a feed-forward layer.
The cross-attention layer is utilized to attend the information from lyrics encoder, , which has been widely applied in previous works on speech synthesis \cite{li2019neural, shen2018natural} and audio generation \cite{makeanaudio, stableaudio}.
For vocal decoder, it models the alignment between the lyrics and vocals.
For accompaniment decoder, it extracts semantic information from the lyrics for generating accompaniment
Moreover, in a complete song, the vocal and accompaniment parts have a complex interrelationship.
The accompaniment must complement the vocal track without overshadowing them, ensuring that both parts work together to highlight the song's expressive and artistic intents.
To understand and model this interrelationship, we introduce a bidirectional cross-attention (BCA) layer, which consists of two symmetrical cross-attention mechanisms.
For example, in the vocal decoder, the BCA allows the model to attend to the generated parts of accompaniment while generating vocals, making arrangements accordingly.
The BCA layer is then defined as follows:
\begin{equation}
    \mathbf{Q}_{v} = \mathbf{H}_{v}\mathbf{W}_{v}^Q,\quad \mathbf{K}_{v} = \mathbf{H}_{a}\mathbf{W}_{v}^K,\quad \mathbf{V}_{v} = \mathbf{H}_{a}\mathbf{W}_{v}^V
\end{equation}
\begin{equation}\label{eq:maskatt}
   \mathbf{M}_{ij}=\begin{cases} 0, \quad \text{allow to attend}\\ -\infty,\quad \text{prevent from attending} \end{cases} 
\end{equation}
\begin{equation} 
    \mathbf{A}_{v} = \text{softmax}\left(\frac{\mathbf{Q}_{v}\mathbf{K}_{v}^\top}{\sqrt{d_k}}+\mathbf{M}\right)
\end{equation}
where $\mathbf{H}_{v}, \mathbf{H}_{a} \in \mathbb{R}^{T \times d_h}$ denote the previous layer's outputs from the vocal decoder and accompaniment decoder, respectively. These outputs are linearly projected to a triple of queries, keys and values with learnable weights $\mathbf{W}_{v}^Q , \mathbf{W}_{v}^K , \mathbf{W}_{v}^V \in \mathbb{R}^{d_h \times d_k}$, respectively, and the mask matrix $\mathbf{M} \in \mathbb{R}^{T \times T} $ is used to control whether a pair of tokens can be attended to each other. Here, we use $T$ to denote the length of tokens in LM, and use $d_h$ and $d_k$ to denote the hidden size and attention layer size.

The vocal decoder and accompaniment decoder treats the generation of semantic tokens as conditional language modeling tasks, performing autoregressive predictions token by token.
Leveraging the in-context learning capabilities of the language model, we can control various acoustic conditions of the generated audio with a prompting technique. 
Given a vocal prompt (represented by semantic tokens), denoted as $\hat{\mathbf{S}}_{v}$, it tends to control a mixture of speaker, vocal melody, and tempo. 
Similarly, given an accompaniment prompt (represented by semantic tokens), denoted as $\hat{\mathbf{S}}_{a}$, it tends to control instruments, musical melody, and rhythm.
The semantic tokens of prompt audio are passed as a prefix to the DSLM and the model uses this prefix to sequentially predict the following token sequence.
Taking the vocal decoder $\boldsymbol\theta_\text{vocal}$ as an example.
The task of the vocal decoder can be formulated as:
\begin{equation}
p(\mathbf{S}_v|\mathbf{C}_\text{lyrics},\hat{\mathbf{S}}_{v};\boldsymbol\theta_\text{vocal}) = \prod_{t=0}^T p(\mathbf{S}_{v,t}|\mathbf{S}_{v,<t},\mathbf{S}_{a,<t},\mathbf{C}_\text{lyrics},\hat{\mathbf{S}}_{v};\boldsymbol\theta_\text{vocal})
\end{equation}

Then, we concatenate the embeddings $\mathbf{E}_v, \mathbf{E}_a\in \mathbb{R}^{T \times d_e}$ from outputs of these two decoders.
The combined embeddings $\mathbf{E}_s\in \mathbb{R}^{T \times 2d_e}$ are fed into a song decoder composed of multiple Transformer blocks to non-autoregressively generate the semantic token sequence for the complete song, achieving a natural and seamless integration of vocals and instruments,
which can be simply represented as:
\begin{equation}
p(\mathbf{S}_s|\mathbf{E}_v,\mathbf{E}_a;\boldsymbol\theta_\text{song}) = \prod_{t=0}^T p(\mathbf{S}_{s,t}|\mathbf{S}_v,\mathbf{S}_a;\boldsymbol\theta_\text{song})
\end{equation}
\subsection{Attention mask strategies for universal song generation}
In both self-attention (SA) layer and bidirectional cross-attention (BCA) layer, we employ the mask matrix $\mathbf{M}$ as shown in Equation \ref{eq:maskatt} to control the access of the semantic tokens to be predicted.
As shown in Figure \ref{fig:dslm}, we implement multiple mask strategies for SA and BCA using different $\mathbf{M}$.

Specifically, we employ two different masking strategies for the SA to control each semantic token's access to the context within the same sequence.
One strategy is the causal attention mask, where the representation of each token can only access the leftward context tokens and itself.
This approach predicts a token conditioned on its historical (left) context, thereby learning generation and continuation capabilities, but it is difficult to fully capture the dependencies between the context.
The other strategy is the non-causal attention mask, where all token can attend to each other within the same sequence.
It incorporates contextual information from the entire sequence, and can generate more comprehensive and enriched context representations than the causal approach.

For BCA, we design four masking strategies to control the mutual attention between the semantic token sequences representing vocals and accompaniment.
The bidirectional mask (BR) allows representations in both the vocal sequence and accompaniment sequence to attend to representations in the other sequence.
However, when predicting the token at time step $t$, it can only attend to the representation of tokens in the other sequence at time step less than or equal to $t$.
For example, the representation $\mathbf{H}_{v,t}$ of semantic token $\mathbf{S}_{v,t}$ 
can only pay attention to $\mathbf{H}_{a,\leq t}$ 
, but not to $\mathbf{H}_{a,>t}$.
It attempts to capture the relationships between vocals and accompaniment, but does not consider the full context of the other sequence, leading to certain limitations when one sequence is pre-determined.
As a supplement, the accompaniment-to-vocals (A2V) and vocals-to-accompaniment (V2A) strategies allow one sequence to attend to all tokens in the other sequence.
Take the A2V as an example, the tokens in vocal sequence can attend to the full context of the accompaniment sequence, while tokens in the accompaniment sequence are not allowed to attend to the vocal sequence.
In this way, the vocal decoder can generate vocals based on the complete accompaniment information.
Similarly, the V2A strategy allows the model to predict accompaniment tokens conditioned on the entire vocals sequence.
Additionally, the None strategy means neither sequence can attend to the other, supporting the independent generation of instrumental music or vocals.

By employing different mask strategies for SA and BCA, as well as the input format, a single SongCreator can achieve competitive performance on multiple song generation tasks, as shown in Table \ref{tab:universal} and Appendix \ref{sec:tasks}.
We also demonstrate in the ablation studies that the specific attention mask we employed for each task are effective.
Furthermore, we support additional tasks shown on our demo page.
\begin{table}
  \scriptsize
  \caption{Specific attention mask strategy of all tasks supported by SongCreator. [$\cdot$] indicates that the condition is optional. * indicates that our proposed model achieves significant improvements in this task.}
  \label{tab:universal}
  \centering
  \begin{tabular}{lcccc}
    \toprule
    \textbf{Tasks} & \textbf{Conditions} & \textbf{Outputs} & \textbf{SA mask} & \textbf{BCA mask}\\
    \midrule
    Lyrics-to-song* & Lyrics, [Vocal prompt], [Accompaniment prompt] & Song, Vocals & Causal, Causal & BR\\
    Lyrics-to-vocals* & Lyrics, [Vocal prompt] & Vocals & Causal, Causal & BR\\
    Accompaniment-to-song & Lyrics, Accompaniment, [Vocal prompt] &  Song, Vocals & Causal, Non-causal & A2V \\
    Vocals-to-song & Vocals, [Lyrics], [Accompaniment prompt] & Song, Music & Non-causal, Causal & V2A\\
    Music continuation & Accompaniment prompt & Music & None, Causal & None \\
    Song editing* & Lyrics, Vocals, Accompaniment& Song, Vocals & Causal, Causal & BR\\
    Vocals editing & Lyrics, Vocals & Vocals & Causal, None & None\\
    Vocals editing in song* & Lyrics, Vocals, Accompaniment& Song, Vocals & Causal, Non-causal & A2V\\
    \bottomrule
  \end{tabular}
\end{table}

\subsection{Training Setup}
We investigate a multi-task training setup, in which the model is trained on several tasks to enhance its composition, arrangement, and comprehension abilities.
We consider the following three tasks:

\paragraph{Song generation from lyrics} In this task, the SA in both the vocals decoder and the accompaniment decoder employs the causal attention mask to simultaneously generate vocal and accompaniment semantic tokens.
For BCA, 80\% of the time we use the bidirectional attention mask to learn how to generate harmoniously coordinated vocals and accompaniment.
In the remaining 20\% of the time, we use the None strategy to allow the model to learn to generate accompaniment or vocal track independently.
This probability setting was inspired by classifier-free guidance related work \cite{le2024voicebox, du2024cosyvoice} to ensure it does not disrupt the training of the BCA.

\paragraph{Song generation from pre-determined accompaniment or vocals} Take the accompaniment is determined as an example, 
in this task, the SA in the vocals decoder maintains the causal mask to generate vocals, while the SA in the accompaniment decoder employs the non-causal mask, with the BCA using the A2V strategy.
Note that for the non-causal mask, we randomly mask 20\% of tokens in the input sequence, to encourage the model to learn the relationships between context tokens.
Furthermore, for the above two training tasks, we provide the model with a vocal and accompaniment prompt to encourage the model to learn to control the acoustic conditions of the generated audio.

\paragraph{Song editing} 
The song editing task combines the above two tasks.
The difference is that we randomly select a span of tokens from the end of the target sequence to replace the audio prompt, using a special token <EDIT> in between to distinguish the editing task from the generation task.

In all training tasks, the vocal decoder and accompaniment decoder are trained using the next token prediction objective, and the song decoder predicts the semantic tokens of the complete song based on the embeddings extracted from the vocal decoder and accompaniment decoder.
After that, we calculate the cross-entropy loss for vocals, accompaniment and song, and optimize the DSLM with the sum of these losses.
Calculating the loss of the song also helps the model effectively reduce the impact of the source separation tool on the overall quality of the generated song.
Note that we follow previous works \cite{aghajanyan2022cm3,le2024voicebox} and calculate the loss on all tokens, not just the masked tokens, for non-causal strategy.
Moreover, we also mask the lyrics 20\% of the time to encourage the model to attempt unconditional generation.

%% file: content/experiments.tex
\section{Experiments}
\subsection{Experimental setup} \label{sec:exp_setup}
\paragraph{Data and model} 
DSLM is trained on 8,500 hours of song data with lyrics (approximately 270,000 songs). 
We employed an automatic speech recognition (ASR) model to provide timestamps for each sentence in the lyrics and a voice activity detection (VAD) model to detect silent segments.
Then, we select appropriate silent segments to split the dataset into 1.7M clips, each no longer than 30 seconds and ensuring the completeness of the sentences.
Each clip is input into the Demucs \cite{rouard2023hybrid,defossez2021hybrid} music source separation model to extract vocals and accompaniment.
Our DSLM has approximately 0.6B parameters.
Detailed configurations are shown in Appendix \ref{sec:dslm_detail}.

\paragraph{Training and Inference}
During training, we train the DSLM for 500K steps using 8 NVIDIA A800 GPUs, with a batch size of 8 for each GPU. Adam optimizer is used with $\beta_1=0.9,\beta_2=0.98,\epsilon=10^{-9}$ and follow the same learning rate schedule in \cite{vaswani2017attention}.
Consistently, top-$k$ sampling is adopted for inference, in which $k$ and temperature are set to 50 and 0.9, respectively.

\paragraph{Evaluations}
Most tasks are evaluated using both objective and subjective metrics.\footnote{Following the setting of DiffSinger \cite{liu2022diffsinger}, vocals generation tasks don’t report the objective results.}
For objective evaluations, Fréchet Audio Distance (FAD) \cite{kilgour19_interspeech} is used to evaluate the generation fidelity; Mel Cepstral Distortion (MCD) is used to measure the spectral distance between the synthesized and Ground Truth; Speaker Embedding Cosine Similarity (SECS) is used for the similarity of speaker identity.
For subjective evaluations, we utilize the commonly used mean opinion score (MOS) tests.
In various tasks, we assess multiple aspects: musicality, quality (focusing on clarity and intelligibility), style similarity (including speaker, melody and instruments), harmony between vocals and accompaniment, and naturalness.
Moreover, AB preference tests are also conducted.
The appendix \ref{sec:exp_detail} shows details of the evaluations.

\paragraph{Baselines}
We conducted comprehensive comparisons between SongCreator and multiple baselines on each task.
First, we establish two baseline models for each task.
One is SongCreator (Single) trained on a specific sequence generation task, and the other replaces DSLM with GPT \cite{gpt2} in SongCreator to predict the target sequence, named GPT.
For lyrics-to-song, we directly conditioned SOTA music generation models, MusicLM \cite{musiclm} and MusicGen \cite{musicgen}, on lyrics to predict songs.
Furthermore, we add another baseline where GPT is used to first predict vocals and then predict the song, named GPT (Vocals \& Song).
For lyrics-to-vocals, in addition to MusicLM, we also introduce the SOTA text-to-speech method VALL-E \cite{valle}.
For vocals-to-song and accompaniment-to-song, we utilize the structure proposed in SingSong \cite{donahue2023singsong} to perform these two tasks, respectively.
To ensure a fair comparison, we replace the semantic and acoustic tokens with BEST-RQ \cite{bestrq} tokens and use our latent diffusion model to convert them into the waveform, establishing another baseline, SingSong (Diffusion).
For music continuation, we employ AudioLM \cite{audiolm} as a baseline.
The detailed implementations of each baseline are shown in Appendix \ref{sec:baseline}.
\subsection{The results of tasks}
\paragraph{Lyrics-to-song}
\begin{table}[ht]
\begin{minipage}{0.50\linewidth}
    \caption{Lyrics-to-song evaluation without audio prompt.}\label{tab:lyrics-to-song}
    \centering
    \resizebox{\linewidth}{!}{
        \begin{tabular}{l|ccc}
        \toprule
        \textbf{Model} & \textbf{FAD} $\downarrow$ &\textbf{Musicality} $\uparrow$ & \textbf{Quality} $\uparrow$\\
        \midrule
        Ground Truth & - & $4.3\pm 0.04$ & $4.09\pm 0.05$\\
        \midrule
        MusicLM & 6.47  & $3.21\pm 0.09$ & $3.25\pm 0.07$\\
        MusicGen & 2.31  & $3.08\pm 0.06$ & $2.99\pm 0.06$\\
        GPT & 8.18 & $3.32\pm 0.10$ & $3.26\pm 0.08$\\
        GPT (Vocals \& Song) & 11.23 & $3.55\pm 0.09$ & $3.64\pm 0.07$\\
        \midrule
        SongCreator & $\mathbf{2.14}$  & $\mathbf{4.25\pm 0.05}$ & $\mathbf{4.08\pm 0.06}$\\
        SongCreator (Single) & 3.04 & $3.85\pm 0.06$ & $3.75\pm 0.05$\\
        \bottomrule
        \end{tabular}}
\end{minipage}
\hfill
\begin{minipage}{0.50\linewidth}
    \caption{Lyrics-to-vocals evaluation without audio prompt.}\label{tab:lyrics-to-vocal}
    \centering
    \resizebox{0.93\linewidth}{!}{
        \begin{tabular}{l|ccc}
        \toprule
        \textbf{Model} &\textbf{Musicality} $\uparrow$& \textbf{Quality} $\uparrow$\\
        \midrule
        Ground Truth & $3.89\pm 0.09$ & $3.91\pm 0.07$\\
        \midrule
        MusicLM & $3.31\pm 0.06$ & $3.35\pm 0.06$\\
        VALL - E & $3.15\pm 0.08$ & $3.23\pm 0.06$\\
        GPT & $3.64\pm 0.07$ & $3.58\pm 0.07$\\
        \midrule
        SongCreator & $\mathbf{3.98\pm 0.04}$ & $\mathbf{3.79\pm 0.05}$\\
        SongCreator (Vocal Only) & $3.68\pm 0.06$ & $3.63\pm 0.05$\\
        SongCreator (Single) & $3.53\pm 0.06$ & $3.64\pm 0.05$\\
        \bottomrule
        \end{tabular}
    }
\end{minipage}
\end{table}
\begin{table}[ht]
\begin{minipage}{0.50\linewidth}
    \caption{Prompt-based lyrics-to-song. We sample the prompt at random from a held-out set.}\label{tab:lyrics-to-song-prompt}
    \centering
    \resizebox{\linewidth}{!}{
        \begin{tabular}{l|cccc}
        \toprule
        \textbf{Model} & \textbf{FAD} $\downarrow$ & \textbf{MCD} $\downarrow$ &\textbf{Musicality} $\uparrow$ & \textbf{Similarity} $\uparrow$\\
        \midrule
        Ground Truth & - & - & $4.04\pm 0.06$ & $3.79\pm 0.09$\\
        \midrule
        MusicGen & $\mathbf{1.90}$ & 9.78 & $3.46\pm 0.11$ & $3.27\pm 0.11$\\
        SongCreator & 2.06 &$\mathbf{8.44}$ & $\mathbf{4.01\pm 0.07}$ & $\mathbf{3.82\pm 0.08}$\\
        \bottomrule
        \end{tabular}
        }
\end{minipage}
\hfill
\begin{minipage}{0.50\linewidth}
    \caption{Prompt-based lyrics-to-vocals. We sample the prompt at random from a held-out set.}\label{tab:lyrics-to-vocals-prompt}
    \centering
    \resizebox{0.90\linewidth}{!}{
        \begin{tabular}{l|cccc}
        \toprule
        \textbf{Model} & \textbf{SECS} $\uparrow$ &\textbf{Musicality} $\uparrow$ & \textbf{Similarity} $\uparrow$\\
        \midrule
        Ground Truth & 0.62 & $3.63\pm 0.08$ & $3.57\pm 0.08$\\
        \midrule
        VALL - E & 0.66 & $3.34\pm 0.07$ & $3.30\pm 0.08$\\
        SongCreator& $\mathbf{0.68}$ & $\mathbf{3.57\pm 0.06}$ & $\mathbf{3.55\pm 0.07}$\\
        \bottomrule
        \end{tabular}
        }
\end{minipage}
\end{table}
As shown in Table \ref{tab:lyrics-to-song}, our proposed SongCreator significantly outperforms the baselines across all three metrics, confirming the effectiveness of SongCreator.
The difference between SongCreator and Ground Truth is merely 0.05 and 0.01 for musicality and quality, respectively.
SongCreator (Single) and GPT (Vocals \& Song) perform better than other baselines, demonstrating the difficulty of directly modeling the complete song.
Additionally, we use the same lyrics from the demos of the previous SOTA model Jukebox \cite{dhariwal2020jukebox} and conduct the AB preference test.
As shown in Table \ref{tab:jukebox}, SongCreator is preferred over Jukebox 60\% of the time.

To investigate the ability of SongCreator to maintain acoustic conditions from prompts, we compared it with MusicGen.
The results are shown in Table \ref{tab:lyrics-to-song-prompt}.
SongCreator achieved scores of 4.01 in musicality and 3.82 in similarity, considerably improving upon MusicGen's scores of 3.46 and 3.27, with only a slightly lower score of 0.16 in FAD.
In addition, SongCreator can independently control the acoustic conditions of the vocals and accompaniment in the generated song.
This capability is lacking in previous methods and results can be found on the demo page.
\paragraph{Lyrics-to-vocals}
SongCreator provides two inference methods for lyrics-to-vocals.
One is similar to lyrics-to-song, where the model considers the relationship between the vocals and accompaniment to generate both vocal and accompaniment tokens, but we only use the generated vocals.
The other doesn't use BCA and the accompaniment decoder, relying solely on the vocal decoder to generate the vocals, named SongCreator (Vocal Only).
As shown in Table \ref{tab:lyrics-to-vocal}, SongCreator (Vocal Only) achieves scores of 3.68 in musicality and 3.63 in quality, comparable to the performance of SongCreator (Single) and GPT.
However, after considering the relationship between vocals and accompaniment, SongCreator surpasses these models with a substantially higher score of 3.98 in musically.
In this study, we also conduct a zero-shot evaluation of the vocals between our proposed model and VALL-E.
Table \ref{tab:lyrics-to-vocals-prompt} presents the results.
From the performance evaluated by MOS and SECS, our proposed model outperforms VALL-E, especially in terms of similarity, demonstrating SongCreator's robust zero-shot clone ability for generating vocals.

\paragraph{Vocals-to-song and accompaniment-to-song}
\begin{table}[b]
\begin{minipage}{0.50\linewidth}
    \caption{Vocals-to-song evaluation.}\label{tab:vocal-to-song}
    \centering
    \resizebox{\linewidth}{!}{
        \begin{tabular}{l|ccc}
        \toprule
        \textbf{Model} & \textbf{FAD} $\downarrow$ &\textbf{Musicality} $\uparrow$ & \textbf{Harmony} $\uparrow$\\
        \midrule
        Ground Truth & -  & $4.12\pm 0.05$ & $3.91\pm 0.08$\\
        \midrule
        SingSong& 3.37  & $3.67\pm 0.10$ & $3.63\pm 0.08$\\
        SingSong (Diffusion)& 4.13 & $3.71\pm 0.08$ & $3.67\pm 0.06$\\
        GPT & 3.07 & $3.73\pm 0.07$ & $3.69\pm 0.07$\\
        \midrule
        SongCreator & 1.88   & $\mathbf{3.77\pm 0.08}$ & $\mathbf{3.77\pm 0.07}$\\
        SongCreator (Single) & $\mathbf{1.46}$ & $3.58\pm 0.08$ & $3.65\pm 0.06$\\
        \bottomrule
        \end{tabular}
        }
\end{minipage}
\hfill
\begin{minipage}{0.50\linewidth}
    \caption{Accompaniment-to-song evaluation.}\label{tab:accompaniment-to-song}
    \centering
    \resizebox{\linewidth}{!}{
        \begin{tabular}{l|ccc}
        \toprule
        \textbf{Model} & \textbf{FAD} $\downarrow$ &\textbf{Musicality} $\uparrow$& \textbf{Harmony} $\uparrow$\\
        \midrule
        Ground Truth & -  & $4.15\pm 0.07$ & $4.11\pm 0.07$\\
        \midrule
        SingSong&1.82  & $3.36\pm 0.06$ & $3.42\pm 0.07$\\
        SingSong (Diffusion)& 2.98  & $3.66\pm 0.06$ & $3.65\pm 0.05$\\
        GPT & 1.64 & $3.53\pm 0.08$ & $3.53\pm 0.09$\\
        \midrule
        SongCreator & 1.24   & $\mathbf{3.67\pm 0.05}$  & $\mathbf{3.78\pm 0.06}$\\
        SongCreator (Single) & $\mathbf{1.23}$  & $3.60\pm 0.07$ & $3.62\pm 0.06$\\
        \bottomrule
        \end{tabular}
    }
\end{minipage}
\end{table}
As shown in Table \ref{tab:vocal-to-song} and Table \ref{tab:accompaniment-to-song}, our proposed SongCreator gets comparable results with recent SOTA models in terms of musicality and harmony.
For the FAD score, our model reaches 1.88 and 1.24 on the two tasks, respectively, outperforming SingSong.
A possible reason is that our model considers the complete song, rather than just the partially separated vocals considered in SingSong.
In addition, we used the same vocals (6 samples) in SingSong's demos to generate songs with our model, and asked subjects to choose their preferred songs.
As shown in Table \ref{tab:singsong}, SingSong gets an extra preference (54.1\%) over SongCreator (30\%).
We speculate one of the reasons is that SingSong uses a large-scale high-quality dataset (46k hours).

\paragraph{Music Continuation}
\begin{wraptable}{r}{6cm}
\vspace{-1em}
\setlength{\tabcolsep}{3pt}
\centering
\caption{Music continuation evaluation.}
\resizebox{1.\linewidth}{!}{
\begin{tabular}{l|ccc}
        \toprule
        \textbf{Model} & \textbf{FAD} $\downarrow$ &\textbf{Musicality} $\uparrow$ & \textbf{Similarity} $\uparrow$ \\
        \midrule
        Ground Truth & -  & $3.9\pm 0.11$ & $3.70\pm 0.10$\\
        \midrule
        AudioLM & 1.33  & $3.95\pm 0.10$ & $3.78\pm 0.08$\\
        GPT & $\mathbf{1.28}$& $3.90\pm 0.10$ & $3.73\pm 0.11$\\
        SongCreator& 1.54  & $\mathbf{3.97\pm 0.08}$ & $\mathbf{3.83\pm 0.08}$\\
        \bottomrule
        \end{tabular}
}
\vspace{-2em}
\label{tab:musiccontinuation}
\end{wraptable}
For the music continuation task, we compare different models by generating 10s music based on a 5s instrumental music prompt. 
As illustrated in Table \ref{tab:musiccontinuation}, we can see that SongCreator achieves comparable results with AudioLM and GPT.
This indicates that SongCreator can effectively continue the musical elements in the prompt, providing the capability to control the accompaniment in song generation.

\paragraph{Editing tasks}
\begin{table}[ht]
\vspace{-1em}
\begin{minipage}{0.53\linewidth}
    \caption{Song editing evaluation.}\label{tab:songediting}
    \centering
    \resizebox{\linewidth}{!}{
        \begin{tabular}{l|cccc}
        \toprule
        \textbf{Model} & \textbf{FAD} $\downarrow$ & \textbf{MCD} $\downarrow$ &\textbf{Musicality} $\uparrow$ & \textbf{Naturalness} $\uparrow$\\
        \midrule
        Ground Truth & - & -  & $4.08\pm 0.07$ & $3.99\pm 0.06$\\
        \midrule
        GPT  & 2.29 & 8.30  & $3.84\pm 0.07$ & $3.72\pm 0.06$\\
        \midrule
        SongCreator & $\mathbf{1.81}$  & 7.90 & $\mathbf{4.01\pm 0.06}$ & $\mathbf{3.78\pm 0.07}$\\
        SongCreator (Single) & 1.87 & $\mathbf{7.85}$ & $3.93\pm 0.08$ & $3.75\pm 0.08$\\
        \bottomrule
        \end{tabular}
        }
\end{minipage}
\hfill
\begin{minipage}{0.47\linewidth}
    \caption{Vocals editing evaluation.}\label{tab:vocalediting}
    \centering
    \resizebox{\linewidth}{!}{
        \begin{tabular}{l|cccc}
        \toprule
        \textbf{Model} & \textbf{SECS} $\uparrow$ &\textbf{Musicality} $\uparrow$ & \textbf{Naturalness} $\uparrow$\\
        \midrule
        Ground Truth  & - & $3.65\pm 0.08$ & $3.45\pm 0.07$\\
        \midrule
        GPT & 0.87  & $3.64\pm 0.07$ & $\mathbf{3.43\pm 0.07}$\\
        \midrule
        SongCreator & 0.87 & $\mathbf{3.68\pm 0.06}$ & $3.31\pm 0.06$\\
        SongCreator (Single) & 0.87 & $3.63\pm 0.06$ & $3.41\pm 0.06$\\
        \bottomrule
        \end{tabular}
        }
\end{minipage}
\end{table}
To evaluate the performance on editing tasks, we manually constructed a dataset of 30 song editing examples, as shown in Appendix \ref{sec:editdata}.
Table \ref{tab:songediting} presents the results of song editing.
We can see that SongCreator gets comparable performance in terms of naturalness to the baselines.
However, benefiting from its strong ability to generate song, SongCreator surpasses these baselines in musicality, achieving a score of 4.01.
In the vocal editing, as shown in Table \ref{tab:vocalediting}, all three models achieve relatively close performance in both subjective and objective evaluations.
To demonstrate the editing ability of SongCreator, we further conduct the AB preference test on three tasks: song editing, vocals editing, and vocals editing in song. 
In each task, SongCreator restores the masked song using its original lyrics and compares it with the audio samples reconstructed using BEST-RQ encoding and LDM decoding to eliminate the potential impact from the encoding and decoding processes during our experiments.
The results are shown in Table \ref{tab:editabs}.
In all tasks, there is no significant difference between the generated song and the Ground Truth ($p > 0.01$), where the p-values are calculated using the Wilcoxon signed-rank test. 
This means that humans judge the edited song produced by SongCreator to be as natural as the original unedited song. 

\subsection{Ablation Studies}
\paragraph{The influence of multi-task training}
Through previous experiments, we can find that multi-task training significant improves most tasks, especially in lyrics-to-song.
This indicates that the DSLM effectively capture the shared information between different tasks, such as composition, arrangement and the relationship between vocals and accompaniment.

\paragraph{The influence of bidirectional cross-attention layer}
\begin{wrapfigure}{r}{0.5\textwidth}
    \centering
    \includegraphics[width=0.48\textwidth]{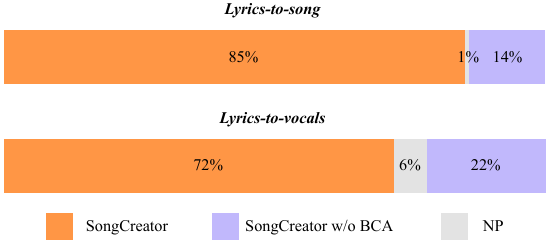}
    \caption{Results of the AB preference test between SongCreator and the model without using BCA.}
    \label{fig:bca}
\vspace{-1.5em}
\end{wrapfigure}
We evaluate the SongCreator and the model without using BCA on lyrics-to-song and lyrics-to-vocals.
Figure \ref{fig:bca} shows the results.
When the BCA is removed from the DSLM, the performance on lyrics-to-song exhibit a marked deterioration, suggesting utilizing BCA is helpful for the model generate harmonious vocals and accompaniment. 
Interestingly, the performance also declined on the lyrics-to-vocals task, demonstrating that learning the relationships between vocals and accompaniment is also beneficial for generating vocals.

\paragraph{The influence of attention mask strategies in self-attention layer}
\begin{wrapfigure}{r}{0.5\textwidth}
\vspace{-1em}
    \centering
    \includegraphics[width=0.48\textwidth]{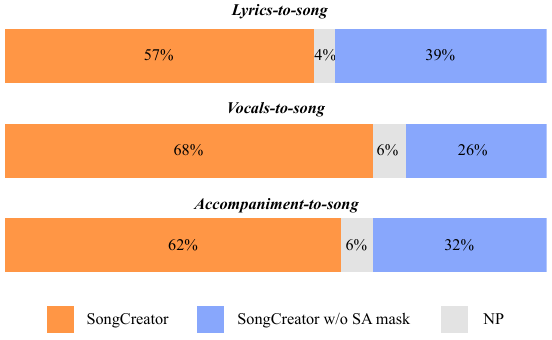}
    \caption{Results of the AB preference test between SongCreator and the model without using non-causal mask in SA.}
    \label{fig:samask}
\vspace{-1.5em}
\end{wrapfigure}
To validate our designed SA mask strategies, we disable the non-causal mask of SA during training and conduct an AB preference test to compare this version with SongCreator on three tasks: lyrics-to-song, vocals-to-song, and accompaniment-to-song.
As shown in Figure \ref{fig:samask}, the performance on all three tasks showed significant degradation, especially for vocals-to-song.
These results indicate that incorporating the non-causal attention mask assists the learning of the relationships within the context and provides additional contextual information for generation.

\paragraph{The influence of attention mask strategies in bidirectional cross-attention layer}
To validate our designed BCA mask strategies, we conduct AB preference tests for the lyrics-to-song and accompaniment-to-song tasks. For lyrics-to-song, we compared BR strategy with A2V, V2A and None strategy. 
As shown in Table \ref{tab:ablationbca1}, replacing the BR strategy with other strategies leads to a significant performance deterioration, demonstrating that the BR strategy is helpful for the model generate harmonious vocals and accompaniment.
The None strategy, which disregards the relationship between vocals and accompaniment, performed the worst. 
In accompaniment-to-song, we compared A2V strategy with BR strategy.
Table \ref{tab:ablationbca2} shows the results, 
We find that participants preferred the song generated with the A2V strategy. 
We believe that this is because the A2V strategy provides more context about the accompaniment sequence when generating vocals.

%% file: content/discussion.tex
\section{Conclusion and Discussion} \label{sec:discussion}
\paragraph{Conclusion} 
In this paper, we propose SongCreator, a system designed for lyrics-based song generation. 
We introduce a dual-sequence language model (DSLM) 
to separately model vocals and accompaniment information,
and employs a dynamic bidirectional cross-attention module to capture the influences between these two sequences, with designing a serious of attention mask strategies for DSLM.
In experiments, the proposed SongCreator provides competitive performance on all eight tasks.
\paragraph{Limitations} 
We acknowledge the limitations of our proposed SongCreator. 
Due to the challenges in collecting data, SongCreator currently cannot control the genre and style of the output songs through text descriptions.
Besides, the interference from accompaniment in the song makes it difficult for BEST-RQ to fully encode the vocal information, imposing a limited clarity of the synthesized vocals -- in further work, we hope to extract better semantic representations for songs.
Another issue is that the proposed model can only generate songs up to 30s, which is insufficient for supporting the generation of songs with complete structures.
\paragraph{Broader Impact}
We believe that our work has huge potential to develop into a song creation tool for content creators or novices to seamlessly express their creative pursuits with a low entry barrier, while also streamline and improve the workflow of experienced music producers.
However, the potential negative impacts of SongCreator can't be overlooked.
One of the primary concerns is the ability to replicate someone's voice with the vocal prompt, which could be exploited in the generation of misinformation, deepfake audio, or any harmful content.
We are committed to advancing the field responsibly, and therefore, the checkpoints trained on the full dataset will not be released.

%% file: content/appendix.tex
\newpage
\appendix
\section{Training and Implementation Details} \label{sec:detail}
\subsection{Dual-sequence language model} \label{sec:dslm_detail}
Our DSLM consists of a lyrics encoder and three decoders.
The lyrics encoder is a 4-layer Transformer \cite{vaswani2017attention} with 1024 hidden size. 
The vocal decoder and accompaniment decoder have a similar architecture that contains 8 DSLM layers with 1024 hidden size.
The song decoder also consists of 4 feed-forward Transformer layers with 1024 hidden size.
We provide detailed hyper-parameter settings about this model configuration in Table \ref{tab:dslm_set}.
We collected approximately 8500 hours of songs with lyrics from the internet for model training, comprising part of the DISCO-10M \cite{lanzendorfer2024disco} dataset and some in-house datasets.

\begin{table}[H]
  \caption{Hyper-parameters of DSLM model}
  \label{tab:dslm_set}
  \centering
  \begin{tabular}{l|l|c}
    \toprule
    \multicolumn{2}{c}{Hyper-parameter}  & Value                 \\
    \midrule
    \multirow{5}*{Lyrics Encoder} & Encoder Layers & 4 \\
    ~ & Hidden Size & 1024 \\
    ~ & Attention Head & 16 \\
    ~ & Feed-Forward Dim & 4096 \\
    ~ & Max Context Length (in \#tokens) & 256 \\
    \midrule
    \multirow{5}*{Vocal Decoder \& Accompaniment Decoder} & Decoder Layers & 8 \\
    ~ & Hidden Size & 1024 \\
    ~ & Attention Head & 16 \\
    ~ & Feed-Forward Dim & 4096 \\
    ~ & Max Context Length (in \#tokens) & 1500 \\
    \midrule
    \multirow{5}*{Song Decoder} & Decoder Layers & 4 \\
    ~ & Hidden Size & 1024 \\
    ~ & Attention Head & 16 \\
    ~ & Feed-Forward Dim & 4096 \\
    ~ & Max Context Length (in \#tokens) & 1500 \\
    \midrule
    \multicolumn{2}{c}{Total Number of Parameters} &  631M \\
    \bottomrule
  \end{tabular}
  \vspace{-0.68em}
\end{table}
\subsection{BEST-RQ with vector quantization} \label{sec:bestrq}
\textbf{BE}RT-based \textbf{S}peech pre-\textbf{T}raining with \textbf{R}andom-projection \textbf{Q}uantizer (BEST-RQ) \cite{bestrq} is a simple and effective self-supervised learning model that learns representations from audio data without manually labeled annotations. 
This self-supervised algorithm helps alleviate the scarcity of song data with lyrics and provides a robust foundation for the entire generation system.

Our implementation of BEST-RQ was based on an open-source library.\footnote{Implemented based on: \href{https://github.com/lucasnewman/best-rq-pytorch}{https://github.com/lucasnewman/best-rq-pytorch}.}
In particular, our implementation follows the same architecture as of BEST-RQ \cite{bestrq}, but with a codebook's vocabulary size of 1024.
For feature extraction, 80-dimensional log Mel-spectrograms are extracted with 24kHz sampling rate with a hop size of 480 and fed into the model to obtain a 50Hz sequence of 1024-dimensional latent representations.
We train this model, which has approximately 0.6 billion parameters, using our prepared 100k hours of audio data in the self-supervised learning (SSL) manner described in \cite{bestrq}.
Furthermore, as we aim to achieve universal song generation, our training dataset includes not only complete songs with vocals and accompaniment but also separate instrumental music and vocals.
This diverse dataset ensures that our model gains a comprehensive understanding of different music elements and their interactions, enhancing its ability to generate a wide array of musical and vocal outputs.

Next, we train a Vector Quantization (VQ) module to quantize the 1024-dimensional latent representations extracted from the 14th layer of the Conformer within the BEST-RQ model.
Our implementation of the VQ module was based on an open-source library,\footnote{Implemented based on: \href{https://github.com/lucidrains/vector-quantize-pytorch}{https://github.com/lucidrains/vector-quantize-pytorch}.} with codebook size of 16384 and codebook dimensional with 32. By combining BEST-RQ and the VQ module, we can extract 50Hz semantic token sequences from the audio.

\begin{table}[H]
  \caption{Reconstructed music performance results for different semantic tokenizers.}
  \label{tab:semantic}
  \centering
  \begin{tabular}{cc}
        \toprule
        \textbf{Model}  &\textbf{ViSQOL}\\
        \midrule
        HuBERT & 2.47\\
        MERT & 2.90\\
        MusicFM & 2.94\\
        \midrule
        BEST-RQ & $\mathbf{3.05}$\\
        \bottomrule
    \end{tabular}
\end{table}

Moreover, we offer a comparison of various prevalent semantic tokenizers, such as HuBERT \cite{hubert}, MERT \cite{mert} and MusicFM \cite{musicfm}.
In this experiment, we train a latent diffusion model (LDM) for each tokenizer to convert sequences of  quantized representations into audio, and randomly selected 50 song segments for fair comparison.
To better evaluate the performance of reconstructing music with different semantic tokenizers, we use ViSQOL \cite{hines2015visqol} as an audio quality assessment metric. 
As shown in Table \ref{tab:semantic}, BEST-RQ providing a greater advantage in music reconstruction.

\subsection{Latent diffusion model} \label{sec:ldm}
As shown in Figure \ref{fig:overall}, we train a latent diffusion model (LDM) as a renderer, which converts a 50Hz semantic token sequence into a 44.1kHz audio, such as songs, vocals, and instrumental music.
In contrast to DDPM \cite{ddpm}, which directly models the raw training data, LDM operates on a low-dimensional latent space to significantly reduce the computational cost and improve the generation stability.
Our implementation of the latent diffusion model was based on the open-source Stable Audio.\footnote{Implemented based on: \href{https://github.com/Stability-AI/stable-audio-tools}{https://github.com/Stability-AI/stable-audio-tools}.}
The reproduced latent diffusion model is composed of a VAE and a U-NET-based conditioned diffusion model.

In particular, for the VAE, we use the same encoder-decoder network architecture as in DAC. \footnote{Implemented based on: 
\href{https://github.com/descriptinc/descript-audio-codec}{https://github.com/descriptinc/descript-audio-codec}.} 
To train the VAE, we first adopted the pre-trained model provided in DAC, then fine-tuned the encoder and decoder components (i.e., replacing the vector quantizers with a diagonal Gaussian re-sampler as in LDM).
We retained the frequency-domain reconstruction loss, discriminators and adversarial loss from DAC and added a KL loss typically used for training VAEs. The VAE was trained on our prepared dataset of 100k hours of songs data, which is the same as the one used for training BEST-RQ.
For the network configurations, the encoder (downsampler) uses strides of [4, 4, 8, 8], d\_model of 128 and latent\_dim of 64, where the 64-dim matrix is employed as the mean and variance of VAE latents (in 32-dim). Besides, the decoder (upsampler) uses strides of [8, 8, 4, 4] and hidden channels of 1536 to transform the 64-dim latents back to 44.1kHz audio. 
Based on this pre-trained VAE, we subsequently train a diffusion model in a way similar to Stable Audio 1.0, except having 32-dim latents as targets and semantic tokens as conditions. 

\section{The details of all tasks supported by SongCreator} \label{sec:tasks}
Benefiting from our specially designed attention mask strategies and multi-task training approach, SongCreator can effectively support the following eight tasks:

\paragraph{Lyrics-to-song} This task aims to generate a complete song that includes harmoniously integrated vocal and accompaniment from lyrics.
Therefore, we use the causal mask for the SA in both vocal and accompaniment decoders to support autoregressive generation.
Regarding the BCA mask strategies, since vocals and accompaniment need to be generated simultaneously, we use the BR strategy to consider the interrelationship between vocals and accompaniment so as to ensure the harmony of vocals and accompaniment.

SongCreator supports to control various acoustic conditions in the generated song by providing optional prompts.
The vocal prompt can control speaker, vocal melody, and tempo, while the accompaniment prompt can control instruments, musical melody, and rhythm.
The vocal prompt and accompaniment prompt can either be present simultaneously, exist individually, or be absent altogether.

\paragraph{Lyrics-to-vocals} This task aims to generate the vocals without accompaniment based on the given lyrics.
As mentioned before, we use the same attention mask strategy as lyrics-to-song task to support better vocal generation.
In this case, the vocal prompt can be provided to control the speaker, melody, and tempo of the generated vocals.

\paragraph{Accompaniment-to-song}
The purpose of this task is to supplement the vocal track of a song based on the given lyrics for a pre-determined accompaniment track.
The vocal track in generated song should complement the input accompaniment track to create a coherent song.
To better encode the contextual representation of the input accompaniment track, we use a non-causal mask strategy for the SA in accompaniment decoder.
For the SA in vocal decoder, we use a causal mask strategy for autoregressive sequence generation.
And to ensure that the generated vocal sequence can consider the full context of the input accompaniment track, we use the A2V strategy in BCA.
Similar to lyrics-to-vocals, the generated vocals can also be controlled using the vocal prompt.

\paragraph{Vocals-to-song}
Contrary to the accompaniment-to-song task, the purpose of this task is to generate harmonious accompaniment for the input vocal track and combine them to create a coherent song.
Thus, in this task, the attention mask strategy is set up in contrast to the Accompaniment-to-song task. 
Similarly, the generated accompaniment can be controlled by the accompaniment prompt.

\paragraph{Music continuation}
This task is expected to generate instrumental music, which is coherent with the accompaniment prompt in terms of instruments, melody, harmony and rhythm.
In this task, we only utilize the accompaniment decoder and use a causal mask in SA for sequence generation.
And to support independent accompaniment sequence generation, we use the None strategy in BCA.

\paragraph{Song editing}
This task requires a model to alter a segment within a song to match a target lyrics.
The modified segment must be coherent with the unedited parts of the original song, i.e., maintaining the speaker, instruments, melody and rhythm.
Considering that it has similar requirements to the lyrics-to-song task, we use the same attention mask strategy.

\paragraph{Vocals editing}
This task is similar to song editing, but the modification target is changed from the complete song to the vocals without accompaniment.
Thus, we only utilize the vocal decoder and use a causal mask in SA for vocal sequence generation.
And to support independent vocal sequence generation, we use the None strategy in BCA.

\paragraph{Vocals editing in song}
This is a unique capability of SongCreator, which modifies the content of the vocal track in a song while keeping the original accompaniment track unchanged.
It means that the modified vocal segment not only maintains coherence with the unedited vocal track of the original song but also harmonizes with the accompaniment in the original song.
Considering that it has similar requirements to the accompaniment-to-song task, we use the same attention mask strategy.

\section{Detailed baseline settings} \label{sec:baseline}
All baselines are trained using similar strategies to those used for DSLM, includes the same dataset, training resources, optimizer settings, and similar parameter scales. Each model was trained for 500K steps. Additionally, for a fair comparison, baselines with semantic tokens as the prediction target (e.g., GPT, SingSong (Diffusion)) shared the same BEST-RQ and LDM modules as DSLM. Here are the implementation details for each baselines:
\paragraph{SongCreator (Single)}
Our proposed SongCreator is trained on multiple tasks. For comparison, we keep the model's structure and hyperparameters and train it on different specific tasks, resulting in SongCreator (Single) for each task.
\paragraph{GPT} 
Inspired by UniAudio \cite{yang2023uniaudio}, we set up this baseline model, treating each task as a conditional language modeling task.
For each task, we first tokenize both the conditional and target audio using BEST-RQ.
Then, we concatenate the source-target pair as a single sequence and perform the next-token prediction task using GPT \cite{gpt2}.
Our implementation of GPT was based on an open-source library,\footnote{Implemented based on: \href{https://github.com/karpathy/nanoGPT}{https://github.com/karpathy/nanoGPT}.} that contains 24 Transformer layers with 1024 hidden size and 4096 feed-forward dimensional.
Finally, we convert the predicted semantic token sequence into audio by the pre-trained latent diffusion model.
\paragraph{MusicLM}
MusicLM \cite{musiclm} has demonstrated excellent performance in text-to-music generation.
Inspired by this, we attempted to employ its methods to lyrics-to-song and lyrics-to-vocals. 
Specifically, to achieve this, we make some modifications to the open-source library.\footnote{Implemented based on: \href{https://github.com/lucidrains/musiclm-pytorch}{https://github.com/lucidrains/musiclm-pytorch}.}
First, we replace the MuLan in MusicLM with a lyrics encoder to better encode phoneme information, and replace w2v-BERT with BEST-RQ to more effectively extract semantic tokens from songs.
Additionally, since SoundStream \cite{zeghidour2021soundstream} is not open-source, we used the widely adopted Encodec \cite{dfossez2023high} as a substitute.
Our reproduced MusicLM follows the same hyperparameters as \cite{musiclm}, using 24 layers of decoder-only Transformers for both the semantic stage and acoustic stage. 
\paragraph{MusicGen}
In addition to MusicLM, MusicGen \cite{musicgen} is another SOTA model in text-to-music generation.
Our implementation of MusicGen for lyrics-to-song is based on the official open-source library.\footnote{Implemented based on: \href{https://github.com/facebookresearch/audiocraft}{https://github.com/facebookresearch/audiocraft}.}
It directly predicts the acoustic tokens extracted by Encodec from the lyrics, without additional semantic tokens.
Similar to other baselines, we also use 24 Transformer layers to ensure this model has approximately 0.6B parameters.
Moreover, considering that MusicGen allows control the generated output through prompts, we also compared it with our proposed SongCreator for the prompt-based lyrics-to-song evaluation.
\paragraph{VALL-E}
Recently, language model-based text-to-speech models (e.g., VALL-E) have shown the capability of generating high-quality personalized speech with a 3s acoustic prompt.
Considering the similarity between text-to-speech and lyrics-to-vocals tasks, we attempted to directly apply VALL-E to the lyrics-to-vocals.
Our implementation is based on the open-source library.\footnote{Implemented based on: \href{https://github.com/lifeiteng/vall-e}{https://github.com/lifeiteng/vall-e}.}
To ensure a fair comparison, both the autoregressive transformer decoder and the non-autoregressive transformer decoder in VALL-E are composed of 24 layers, 16 attention heads, an embedding dimension of 1024, and feed-forward layers of dimensionality 4096.
And we also compared the zero-shot voice cloning abilities of SongCreator and VALL-E.
\paragraph{AudioLM}
To validate the  performance of SongCreator in music continuation, we implement AudioLM \cite{audiolm} based on the open-source code.\footnote{Implemented based on: \href{https://github.com/lucidrains/audiolm-pytorch}{https://github.com/lucidrains/audiolm-pytorch}.}
Similar to our settings with MusicLM, we replace w2v-BERT with BEST-RQ and Soundstream with Encodec in AudioLM.
Additionally, we also used a 24-layer decoder-only transformer structure for both the semantic and acoustic stages.
\paragraph{SingSong}
SingSong \cite{donahue2023singsong} has demonstrated excellent performance in vocals-to-accompaniment generation.
In this work, we reproduce SingSong based on our previous implementation of AudioLM and utilize it as a baseline for the vocals-to-song task.
We follow the same setup as SingSong \cite{donahue2023singsong}, which generates the accompaniment based on the vocals first and then mixes the vocals and accompaniment to produce the complete song.
To eliminate the influence of the pre-trained latent diffusion model, we also directly use it to convert the semantic tokens predicted by SingSong into audio without requiring an additional acoustic modeling stage. 
This new baseline is named SingSong (BEST-RQ).

Additionally, we established two baselines for the accompaniment-to-song task by concatenating the lyrics as another condition before the semantic token sequence of accompaniment. 
In this setup, the prediction target of this model is the semantic tokens of vocals  in the song.

\section{The editing dataset} \label{sec:editdata}
We performed insertion, deletion or substitution operations on the original lyrics, with editing spans ranging from 1 to 15 words.
Examples of the song editing dataset are shown in table \ref{tab:editdata}.
\begin{table}[H]
  \caption{Examples of the song editing dataset.}
  \label{tab:editdata}
  \centering
  \begin{tabular}{cp{5.2cm}p{5.2cm}}
    \toprule
    Edit Types & Original Lyrics  & Edited Lyrics \\
    \midrule
    insertion & Will you let me know. If you have the key. & Will you let me know. If you \textbf{truly hold the answer to my heart,} have the key.\\
    deletion & Was it something that i said? I tried my best, yeah, all for you. \textbf{I can see it in your eyes,} the way you lie, I'm such a fool; & Was it something that I said? I tried my best, yeah, all for you. The way you lie, I'm such a fool.  \\
    substitution &  Cause you had your chance, but \textbf{he chose her first. It must be time, it must be time} that heals. You must need time. & Cause you had your chance, but \textbf{she caught his eye before you. Seems like fate, seems like fate} that heals. you must need time. \\
    substitution & \textbf{I can remember the feeling of being small. Praying} to a god I don't believe in. & \textbf{Feeling so tiny, whispering} to a god I don't believe in.\\
    substitution & Love you forever, that's for sure. \textbf{And promise, I'll keep you warm.} Our love, shines bright like the sun. & Love you forever, that's for sure. \textbf{Vow to hold you close, provide comfort through every storm, warm.} Our love, shines bright like the sun.\\
    deletion & I don't wanna be adored I wanna adore you. \textbf{You're in a car driving home.} Leaving oceans and mountains between us, oh no. & I don't wanna be adored I wanna adore you. Leaving oceans and mountains between us, oh no.\\
    insertion & Aa-aa-ah. Losing all emotions, I didn't feel nothing. In need of attention but nobody's looking. & Aa-aa-ah. Losing all emotions, I didn't feel nothing. in need of \textbf{a sign, a glance, just something to show me I'm seen} of attention but nobody's looking.\\
    substitution & I'm take a sad girl, turn into a \textbf{bad girl}. Even at your worst, you're better than the rest. & I'm take a sad girl, turn into a \textbf{queen}. Even at your worst, you're better than the rest.\\
    \bottomrule
  \end{tabular}
\end{table}

\section{Results of the AB preference test}
\begin{table}[H]
  \caption{Results of the AB preference test between SongCreator and Jukebox in lyrics-to-song. N/P denotes ``no preference".}
  \label{tab:jukebox}
  \centering
  \begin{tabular}{ccc}
        \toprule
        \textbf{SongCreator}  &\textbf{Jukebox} & \textbf{N/P}\\
        \midrule
        60\% & 38.5\% & 1.5\%\\
        \bottomrule
    \end{tabular}
\end{table}
\begin{table}[H]
  \caption{Results of the AB preference test between SongCreator and Singsong in vocals-to-song. N/P denotes ``no preference".}
  \label{tab:singsong}
  \centering
  \begin{tabular}{ccc}
        \toprule
        \textbf{SongCreator}  &\textbf{SingSong} & \textbf{N/P}\\
        \midrule
        30\% & 54.1\% & 15.9\%\\
        \bottomrule
    \end{tabular}
\end{table}
\begin{table}[H]
  \caption{Results of the AB preference test between SongCreator and the Ground Truth in different edit tasks. N/P denotes ``no preference".}
  \label{tab:editabs}
  \centering
  \begin{tabular}{l|ccc}
        \toprule
        \textbf{Tasks} & \textbf{SongCreator}  &\textbf{Ground Truth} & \textbf{N/P}\\
        \midrule
        Song Editing & 48\% & 40\% & 12\%\\
        Vocal Editing & 53\% & 33\% & 14\%\\
        Vocal Editing in Song & 32\% & 52\% & 16\%\\
        \bottomrule
    \end{tabular}
\end{table}
\begin{table}[H]
  \caption{Results of the AB preference test for using different attention mask strategies in BAC on the lyrics-to-song task.}
  \label{tab:ablationbca1}
  \centering
  \begin{tabular}{ccccc}
        \toprule
        \textbf{BR} & \textbf{A2V}  &\textbf{V2A} & \textbf{None} & \textbf{N/P}\\
        \midrule
        76\% & 20\% & - & -  & 4\%\\
        71\% & - & 25\% & -  & 4\%\\
        85\% & - & - & 14\%  & 1\%\\
        \bottomrule
    \end{tabular}
\end{table}

\begin{table}[H]
  \caption{Results of the AB preference test for using different attention mask strategies in BAC on the Accompaniment-to-song task.}
  \label{tab:ablationbca2}
  \centering
  \begin{tabular}{ccc}
        \toprule
        \textbf{BR} & \textbf{A2V}  & \textbf{N/P}\\
        \midrule
        27\% & 59\% & 14\%\\
        \bottomrule
    \end{tabular}
\end{table}

\section{Results of the inference speed}
\begin{table}[H]
  \caption{The real-time factor (RTF) for different models on a single NVIDIA V100 GPU with a batch size of 1 during inference. }
  \label{tab:rtf}
  \centering
  \begin{tabular}{l|c}
        \toprule
        \textbf{Model} & \textbf{RTF} \\
        \midrule
        MusicLM & 14.545\\
        MusicGen & 2.104\\
        GPT & 1.525\\
        GPT (Vocals \& Song) & 3.059\\
        \midrule
        SongCreator & 2.793\\
        \bottomrule
    \end{tabular}
\end{table}
To evaluate the inference speed, we supplement the evaluation by comparing the real-time factor (RTF) for SongCreator and other baselines. RTF represents the time (in seconds) required for the system to synthesize one second of waveform. The evaluation was performed on a single NVIDIA V100 GPU with a batch size of 1. We randomly selected 20 generated audio samples, each longer than 20 seconds, to conduct the evaluation. 

As shown in Table \ref{tab:rtf}, the results indicate that methods utilizing a single LM module are significantly faster than MusicLM, which employs multiple LMs in cascading manner. Taking into account the experiments corresponding to Table \ref{tab:lyrics-to-song}, we observe that although GPT and MusicGen, which only model the song token sequence, are faster than GPT (Vocals \& Song) and SongCreator, which predict multiple sequences, this gain in speed comes at the cost of reduced performance. In comparison to GPT (Vocals \& Song), our proposed SongCreator, which leverages DSLM to simultaneously model both vocals and accompaniment, achieves not only faster speeds but also better results.

\section{Detailed experimental settings} \label{sec:exp_detail}
\subsection{Details in objective evaluations}
Here, we provide details of the objective evaluations.
\paragraph{FAD}
Fréchet Audio Distance (FAD) \cite{kilgour19_interspeech} is used to evaluate the generation fidelity of music.
We calculate FAD based on the distribution distance between the feature of the target an generated audios, extracted from VGGish \cite{hershey2017cnn} model.
\paragraph{MCD}
Mel-cepstral distortion (MCD) is a signal-level quality metrics derived from human auditory research, which measures the spectral distance between the synthesized and reference mel-spectrum features.
In our research, we attempt to use it to indicate the distance between the generated song and the Ground Truth.
\paragraph{SECS}
Speaker Embedding Cosine Similarity (SECS) is a widely used metrics in the speech generation, employed to evaluate the similarity of speaker identify.
We use the speaker encoder of the Resemblyzer package\footnote{Implemented based on: \href{https://github.com/resemble-ai/Resemblyzer}{https://github.com/resemble-ai/Resemblyzer}} to compute the SECS between
the prompt vocals and synthesized vocals.
\subsection{Details in subjective evaluations} \label{sec:mos_detail}
For lyrics-to-song and lyrics-to-vocals, we focus on the musicality and quality of the generated songs. 
We conducted MOS (Mean Opinion Score) tests for both aspects, providing subjects with detailed descriptions, and report both mean and CI95 scores of our MOS tests.
In these tests, subjects are specifically asked to focus on the musicality and quality of the song in each respective test.
The subjects present and rate the samples, and each subject is asked to evaluate the subjective musicality and quality on a 1-5 scale.

For the prompt-based lyrics-to-song, prompt-based lyrics-to-vocals and music continuation, in addition to musicality, we also asked subjects to focus on the similarity between the generated vocals and accompaniment (if present) to the provided reference audio. 
In this evaluation, subjects are instructed to ignored the differences in content and audio quality, and to evaluate how well the synthesized results matched the reference audio.

For the prompt-based vocals-to-song and accompaniment-to-song, in addition to musicality, we follow SingSong \cite{donahue2023singsong} to ask subjects to focus on the harmony between the vocals and accompaniment. 
We write explicit instructions to ask the subjects to assess the generated song.

For song editing and vocals editing, in addition to musicality, we also conduct naturalness MOS.
This test is aimed to make subjects judge whether the audio appears to have been edited based on its naturalness.
In addition,  AB preference test is also conducted to ask subjects to give their preferences between a pair of songs.

\begin{figure}
     \centering
     \includegraphics[width=\textwidth]{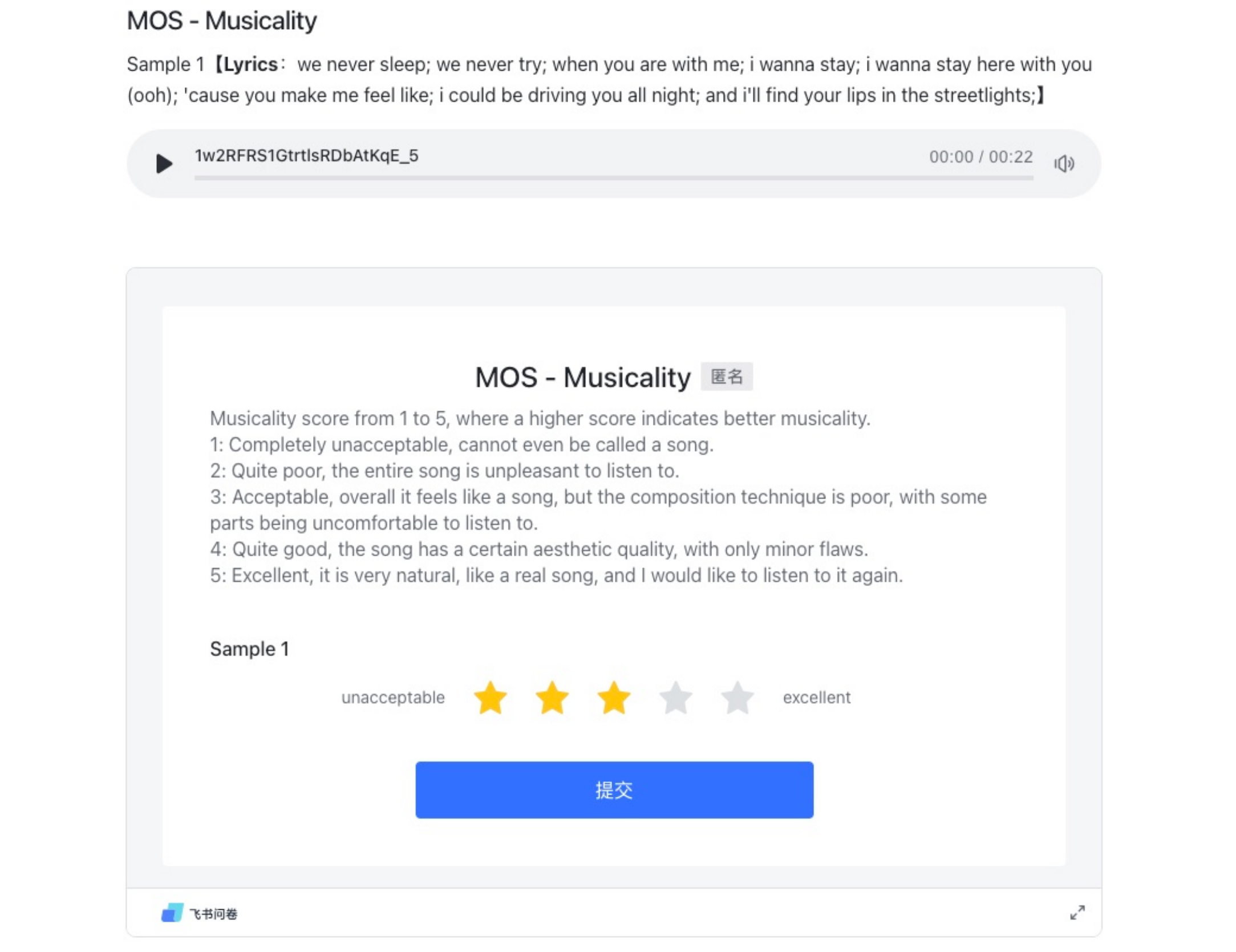}
     \caption{The screenshot of MOS test in musicality evaluation.}
     \label{fig:mos-musicality}
\end{figure}

\begin{figure}
     \centering
     \includegraphics[width=\textwidth]{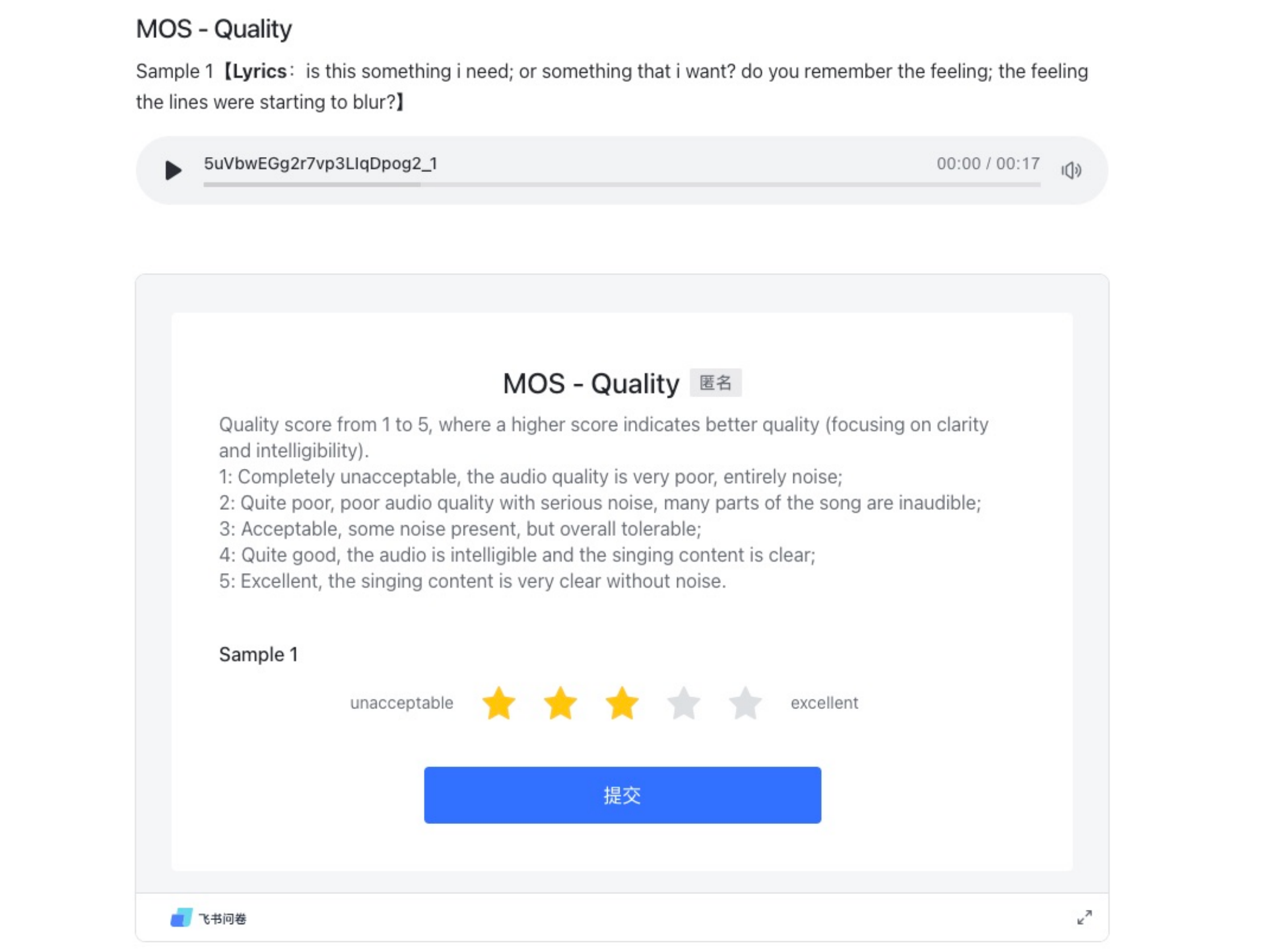}
     \caption{The screenshot of MOS test in sound quality evaluation.}
     \label{fig:mos-quality}
\end{figure}

\begin{figure}
     \centering
     \includegraphics[width=\textwidth]{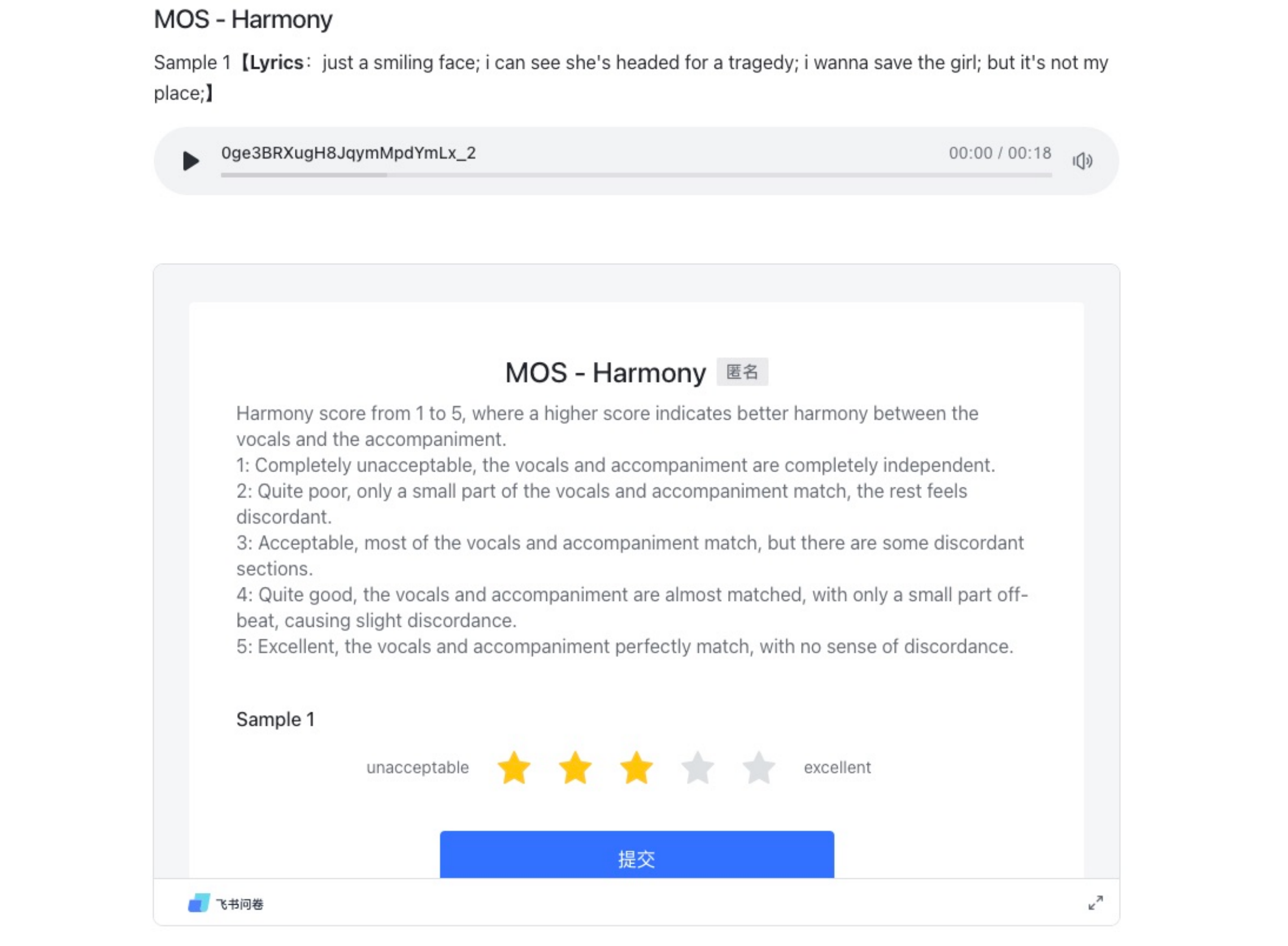}
     \caption{The screenshot of MOS test in harmony evaluation.}
     \label{fig:mos-harmony}
\end{figure}

\begin{figure}
     \centering
     \includegraphics[width=\textwidth]{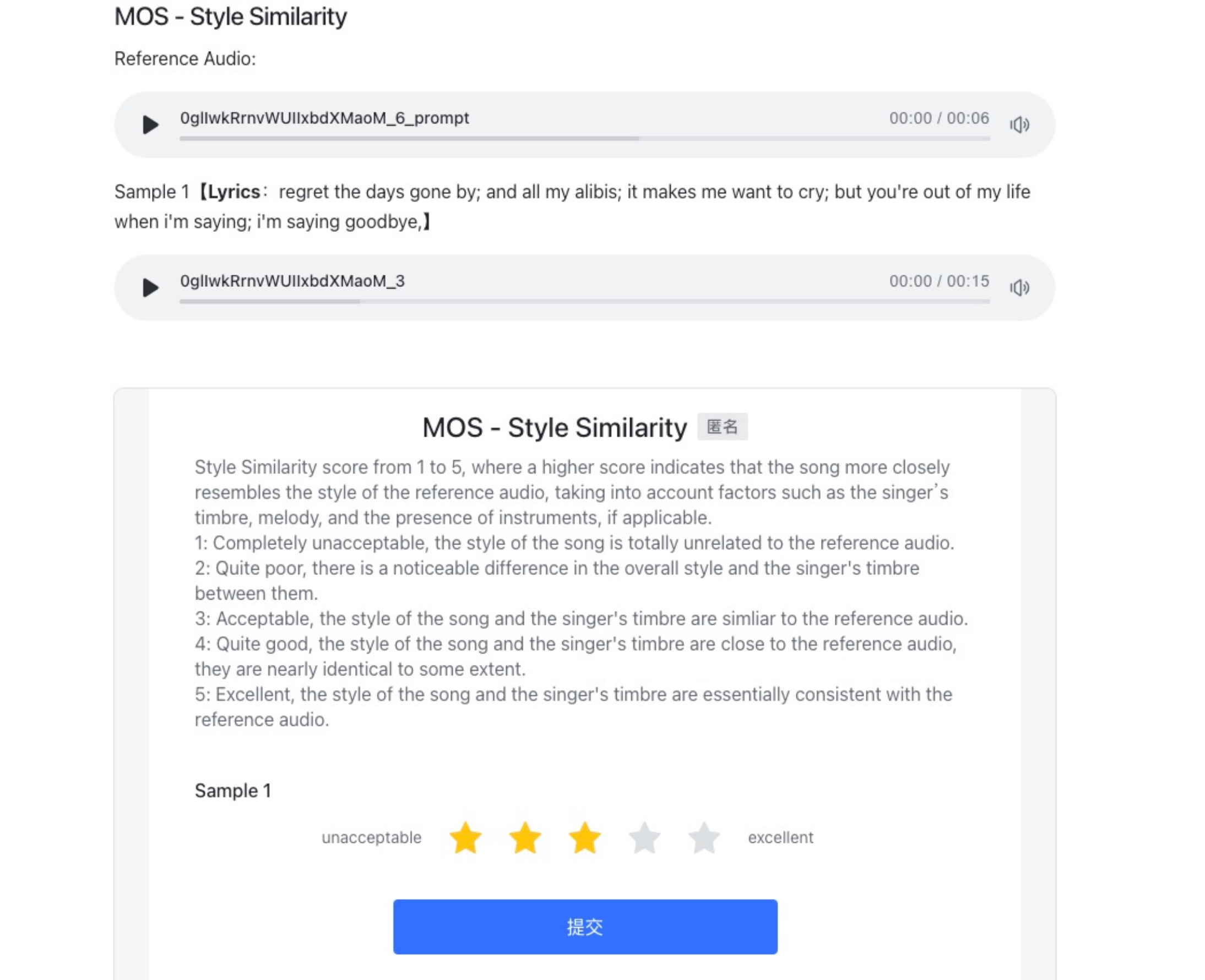}
     \caption{The screenshot of MOS test in similarity evaluation.}
     \label{fig:mos-similarity}
\end{figure}

\begin{figure}
     \centering
     \includegraphics[width=\textwidth]{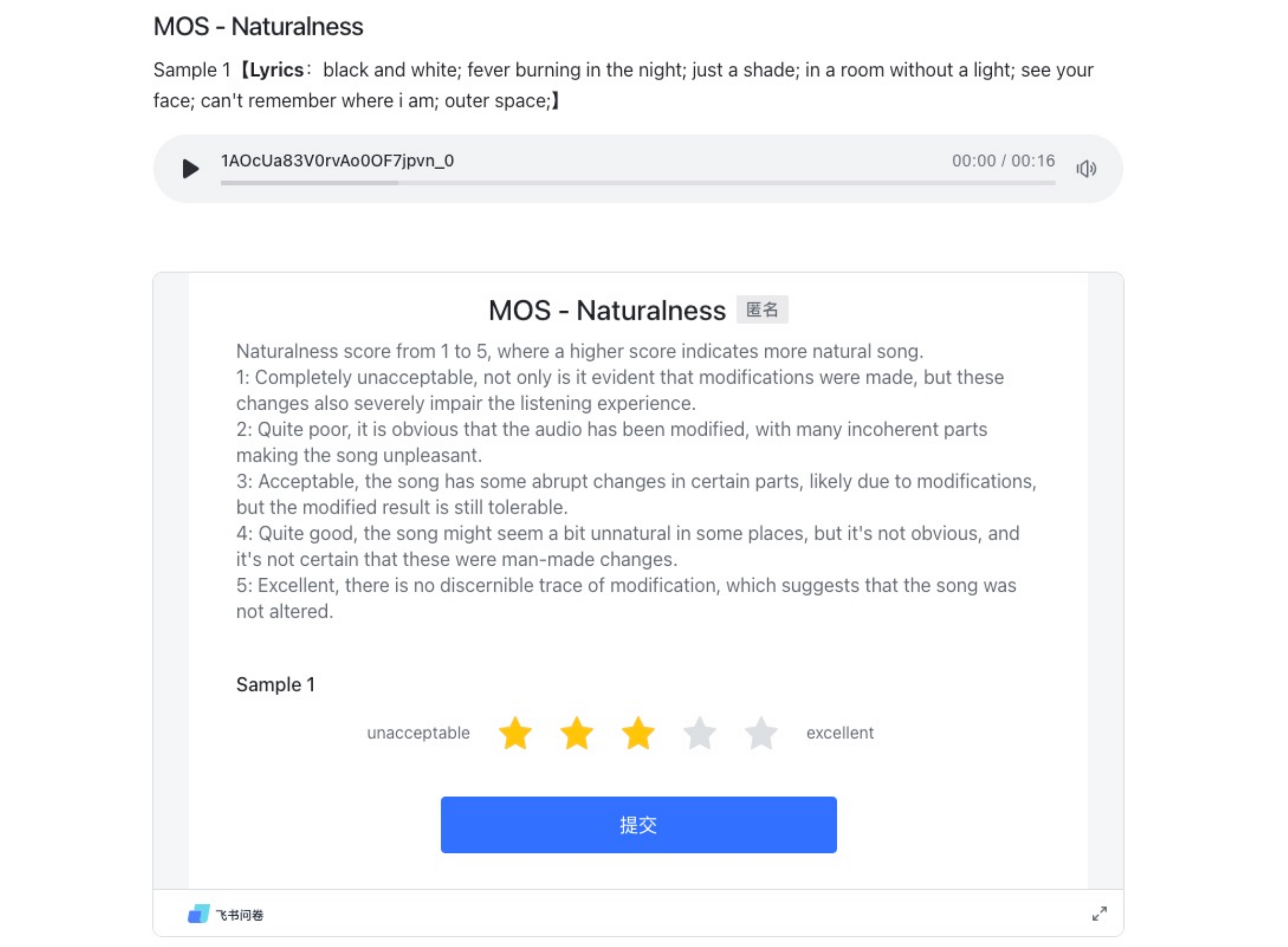}
     \caption{The screenshot of MOS test in naturalness evaluation.}
     \label{fig:mos-naturalness}
\end{figure}

\begin{figure}
     \centering
     \includegraphics[width=\textwidth]{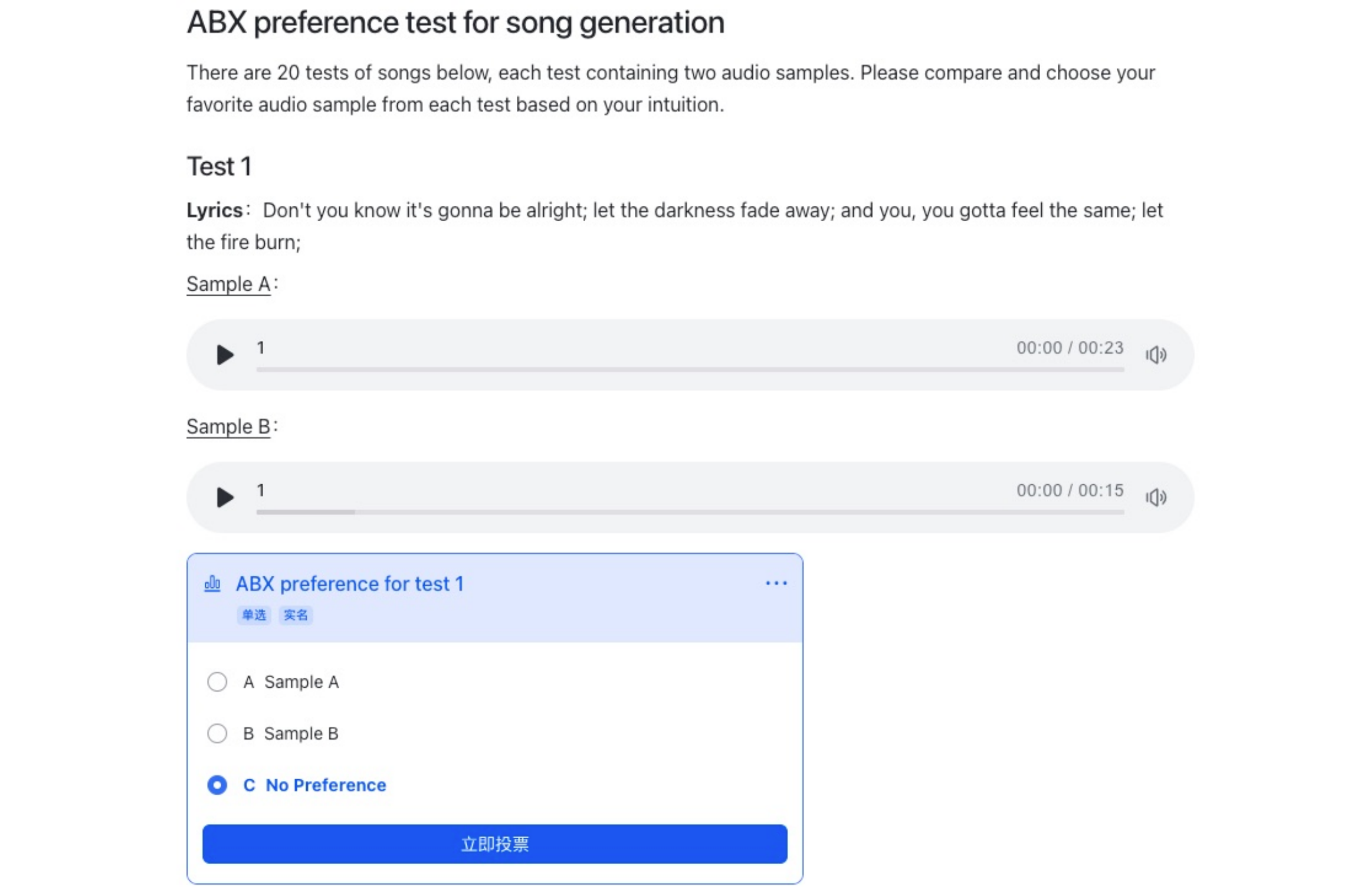}
     \caption{The screenshot of AB preference test.}
     \label{fig:screenshot_abx_1}
\end{figure}

\begin{figure}
     \centering
     \includegraphics[width=\textwidth]{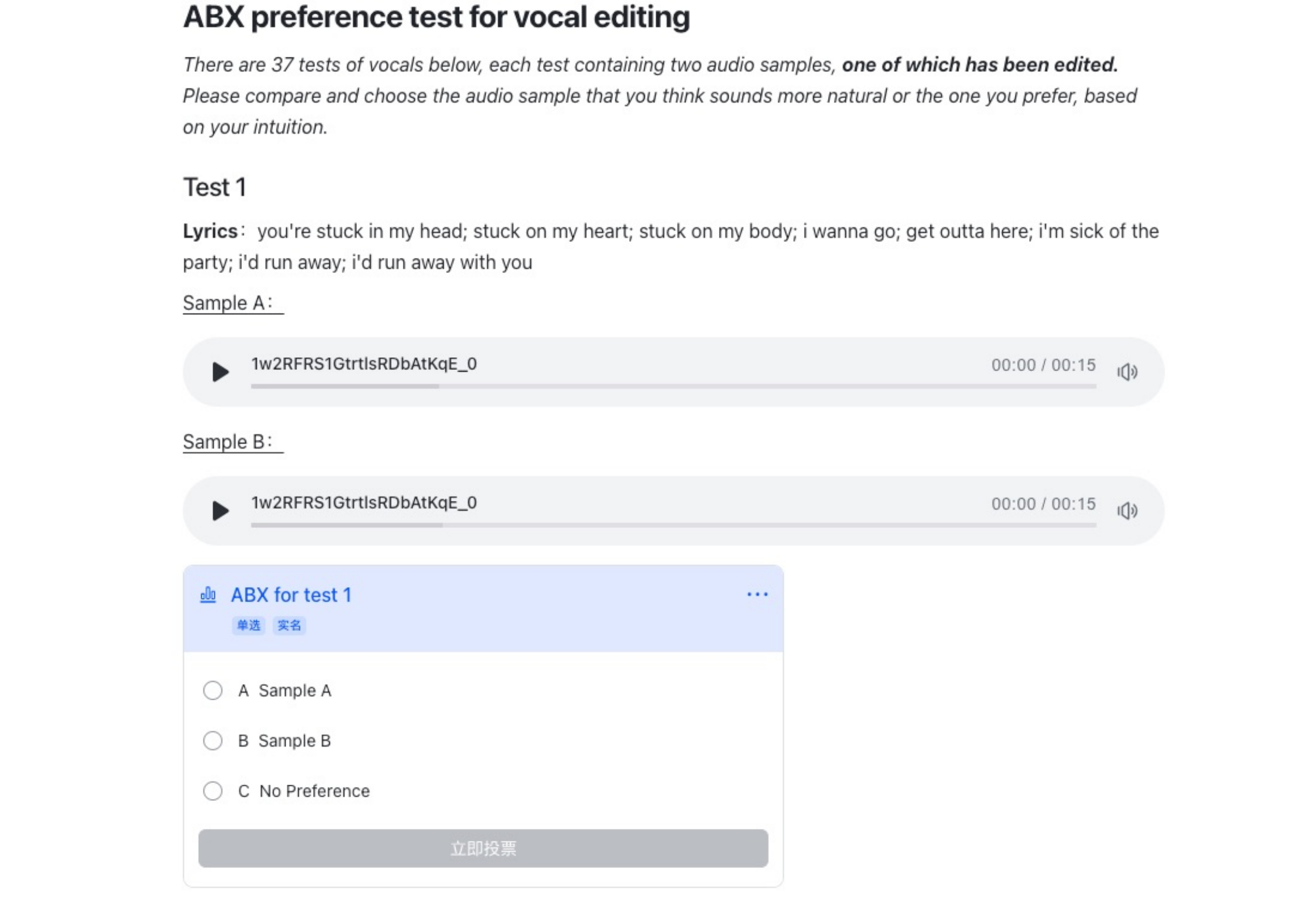}
     \caption{The screenshot of AB preference test.}
     \label{fig:screenshot_abx_2}
\end{figure}

Our MOS tests are crowd-sourced and conducted by 25 listening subjects, while the AB preference tests are conducted by 20 listening subjects.
All the screenshots of instruction for subjects have been shown in 
Figure \ref{fig:mos-musicality}-\ref{fig:screenshot_abx_2}.
We paid \$$10$ to subjects hourly and totally spent about \$$600$ on participant compensation.
We tell the subjects that the data will be used in scientific research.

